\documentclass[12pt]{article}
\usepackage{mathtools, bm, graphicx, hyperref, mathrsfs, bbm}
\usepackage{amsfonts, amsmath, amssymb, amssymb, amsthm}
\usepackage{dsfont}
\usepackage{color}
\usepackage[dvipsnames]{xcolor}
\usepackage[all, knot]{xy}
\usepackage{tikz}
\usepackage{braket}
\usepackage{epstopdf}
\usepackage[footnotesize]{caption}
\usepackage{enumitem}   
\usepackage{subeqnarray}         
\usepackage{cases}               
\usepackage{subfigure}
\usepackage{cite}               
\usepackage{hyperref}            
\usepackage{multirow,makecell}   
\usepackage{textcomp}
\usepackage{wasysym}
\usepackage{url}
\usepackage[margin=3cm]{geometry}
\usepackage{geometry}
\usepackage{float}
\usepackage{tensor}
\usepackage{appendix}
\allowdisplaybreaks[3]

\begin{document}
\begin{titlepage}
\vspace{0.5cm}
\begin{center}
{\Large\bf{Note on holographic torus stress tensor correlators in $AdS_3$ gravity}}
\lineskip .75em
\vskip 2.5cm
{\large{Song He$^{\clubsuit,\maltese,}$\footnote{hesong@jlu.edu.cn}, Yi Li$^{\clubsuit,}$\footnote{liyi@fudan.edu.cn}, Yun-Ze Li$^{\clubsuit,}$\footnote{lyz21@mails.jlu.edu.cn}, Yunda Zhang$^{\clubsuit,}$\footnote{ydzhang21@mails.jlu.edu.cn}}}
\vskip 2.5em
{\normalsize\it $^\clubsuit$Center for Theoretical Physics and College of Physics, Jilin University,\\
 Changchun 130012, People's Republic of China
\\$^{\maltese}$Max Planck Institute for Gravitational Physics (Albert Einstein Institute),\\ 
Am M\"uhlenberg 1, 14476 Golm, Germany}
\end{center}
\begin{abstract}
In the AdS$_3$/CFT$_2$ framework, the Euclidean BTZ black hole corresponds to the dominant high-temperature phase of its dual field theory. We initially employ perturbative methods to solve the Einstein equations as boundary value problems, providing correlators for the energy-momentum tensor operator at low points. Utilizing operator equations established in our previous work, we further compute arbitrary high-point correlators for the energy-momentum tensor operator in the high-temperature phase and recursive relations for these high-point functions.  Concurrently, we employ the Chern-Simons formalism to derive consistent results. Further, using the cut-off AdS/$T\bar{T}$-deformed CFT duality, we calculate the energy-momentum tensor correlators, contributing to the comprehensive understanding of the system's dynamics.  Finally, stress tensor correlators enable us to ascertain the corresponding KdV operator correlators at low-temperature. 
\end{abstract}
\end{titlepage}

\newpage
\tableofcontents

\section{Introduction}\label{1}

The AdS/CFT correspondence \cite{Maldacena:1997re,Gubser:1998bc,Witten:1998qj} is an statement of equivalence between a quantum gravity theory in the $d+1$ dimensional asymptotically AdS bulk space and a conformal field theory on the $d$ dimensional conformal boundary. The dictionary of the correspondence is the Gubser-Klebanov-Polyakov-Witten (GKPW) relation \cite{Gubser:1998bc,Witten:1998qj} that the gravitational partition function with prescribed boundary condition is equal to the partition function with operator insertion of the conformal field theory
\begin{align} \label{GKPW relation}
 Z_{G}[\psi^{(0)}] = Z_{CFT}[\psi^{(0)}]
\end{align}
$\psi^{(0)}$, collectively representing boundary conditions of all bulk fields on the gravity side, are sources of the dual operators on the conformal field theory side. Therefore, correlators in a CFT correspond to scattering amplitudes, or holographic correlators, in the dual $\text{AdS}$ gravity. In the large central charge ($c$) limit of CFT/semiclassical limit of gravity, the gravity partition function can be approximated by saddle points, and computation of holographic correlators reduces to a problem of solving classical equations of motion with prescribed boundary conditions.

We consider torus correlators of the stress tensor in $\text{AdS}_3/\text{CFT}_2$ in this paper. There has been extensive study of CFT on a Riemann surface, in particular, correlators of stress tensors were studied in \cite{Friedan:1986ua,Eguchi:1986sb,He:2022jyt}, where insertion of the stress tensor operator was related to taking derivative with respect to the moduli. On the holography side, see \cite{Carlip:2005zn,Kraus:2006wn} for a review of early works in $\text{AdS}_3/\text{CFT}_2$. The holographic torus partition function was studied in \cite{Maloney:2007ud} from the Einstein-Hilbert theory of pure 3D gravity, higher genus cases were studied in \cite{Yin:2007gv} and a large $c$ CFT computation of the partition function was later given in \cite{Chen:2015uga}. Recently, a top-down derivation of $\text{AdS}_3/\text{CFT}_2$ from string theory was done in \cite{Eberhardt:2019ywk}. In principle, if the partition function is given as a function on the moduli space as in \cite{Maloney:2007ud,Yin:2007gv}, stress tensor correlators can be computed according to \cite{Friedan:1986ua,Eguchi:1986sb} by differentiating the partition function with respect to the moduli. Following this spirit, we derive an explicit recurrence relation of torus correlators of the stress tensor, and all higher point correlators can be obtained from the one-point correlator or the partition function computed holographically from classical gravity saddles. Alternatively, we show the stress tensor correlators can also be computed directly from the dynamics of the Einstein-Hilbert theory of pure 3D gravity, by solving the boundary value problem of Einstein's equation. There has been a thorough study on the local (near conformal boundary) solution of this boundary value problem \cite{Fefferman:1985ci,Henningson:1998gx, Skenderis:1999nb,deHaro:2000vlm,Fefferman:2007rka} in the form of the Fefferman-Graham series, and it successfully recovers holographic UV properties of the CFT, e.g., the Weyl anomaly. On the other hand, the problem of global solution is more difficult \cite{Graham:1991jqw,Biquard:2000me,anderson2004structure,anderson2008einstein,Anderson:2004yi}, and explicit expressions of holographic stress tensor correlators are only known in the case of pure $AdS_{d+1}$ \cite{Liu:1998ty,DHoker:1999bve,Arutyunov:1999nw,Raju:2012zs,Nguyen:2021pdz}, where the symmetry significantly simplifies the problem. Our computation of holographic torus correlators of stress tensor is an example of solving the boundary value problem in a slightly less trivial geometry. Holographic stress tensor correlators have already been also considered in the AdS Schwarzschild black holes, (in the Minkowski signature thus a different kind of boundary value problem) to compute holographic transport coefficient in \cite{Policastro:2001yc} and many following works, and to compute holographic OPE coefficients in \cite{Karlsson:2022osn}. In both cases, only part of the information of the correlators has been worked out and the full boundary value problem remains to be solved.

This program of computing holographic stress tensor correlators can be extended to the context of cutoff AdS/$T\bar{T}$ CFT holography. It was proposed that for $T\bar{T}$ deformation \cite{Zamolodchikov:2004ce,Smirnov:2016lqw} of holographic CFTs, the holographic dual is to move the conformal boundary to a finite cutoff in the bulk AdS space \cite{McGough:2016lol}. Following this spirit, stress tensor correlators in the large $c$ limit in the complex plane and holographic correlators on a cutoff plane in $\rm AdS_3$ were computed in \cite{Kraus:2018xrn,Aharony:2018vux,Li:2020pwa}. Another equivalent holographic dual of $T\bar T$ deformation is to impose a mixed boundary condition at infinity \cite{Guica:2019nzm}. In this description, the on-shall action \cite{Tian:2023fgf,Tian:2024vln} and the one-loop partition function \cite{He:2024pbp} have been obtained. In our previous research \cite{He:2023hoj}, we investigated the holographic correlators on a cutoff torus in thermal AdS$_3$, corresponding to the low-temperature phase. As a natural extension, in this paper, we compute the holographic torus correlators in the cutoff BTZ black hole, corresponding to the high-temperature phase.

The remainder of this manuscript is organized as follows. We start by computing the holographic torus correlators of stress tensor on the conformal boundary in Section 2. In Subsection 2.1, we compute the correlators by directly solving Einstein's equation on the Euclidean BTZ saddle. In Subsection 2.2, we derive a recurrence relation by relating the insertion of a stress tensor operator to differentiation with respect to the modular parameter of the torus. In Subsection 2.3, we discuss the contribution to the correlator from other saddles. In Section 3, we apply the Chern-Simons formalism of 3D gravity to evaluate the stress tensor correlators. In Section 4, we apply the cutoff AdS/$T\bar{T}$ CFT correspondence to investigate the holographic torus stress tensor correlators in cutoff BTZ black hole. With stress tensor correlators, the correlation function of KdV operators can be obtained. Finally, we end in section 6 with a summary and an outlook.

\section{Holographic correlators of stress tensor on the conformal boundary}\label{2}
In this section, we compute the holographic torus correlators of the stress tensor on the conformal boundary in three-dimensional gravity of the minimal Einstein-Hilbert theory. We start by reviewing the basics of holographic correlators.

Restricting ourselves to correlators of the stress tensor, sources/bulk fields other than the background metric/bulk metric are turned off in (\ref{GKPW relation}). If one classical gravity saddle dominates, the generating functional (of connected correlators) of the holographic conformal field theory with background metric $\gamma$ is equal to the on-shell action of the saddle with $\gamma$ being the boundary condition for its bulk metric,
\begin{equation}
	I_{CFT}[\gamma] = I_{G-on shell}[\gamma].
\end{equation}
The correlators \footnote{Unless otherwise specified, correlator meanings connected correlator throughout this work.} of the stress tensor are obtained by taking the functional derivative with respect to the metric, evaluated in the background geometry,
\begin{align} \label{correlator from generating functional}
    \langle T_{i_1j_1}(z_1) \ldots T_{i_nj_n}(z_n)\rangle = -\frac{(-2)^n \delta^n}{\sqrt{\det(\gamma(z_1))}\ldots \sqrt{\det(\gamma(z_n))}\delta \gamma^{i_1j_1}(z_1) \ldots \delta\gamma^{i_nj_n}(z_n)} I_{CFT}.
\end{align}
To account for the IR divergence of the gravity on-shell action in an asymptotic AdS Einstein space, we need to regulate by introducing a cutoff surface and local counter terms to cancel the divergence and take the limit that the cutoff surface goes to the conformal boundary. The action of Einstein-Hilbert theory of pure gravity, with the Gibbons-Hawking term \cite{York:1972sj, Gibbons:1976ue} and the proper counter term \cite{Balasubramanian:1999re, Emparan:1999pm,deHaro:2000vlm}, is\footnote{The AdS radius $l$ is set to 1.}
\begin{equation}
    I_G=-\frac{1}{16\pi G}\int_{\mathcal{M}}d^3x\sqrt{\det(g)}(R-2)-\frac{1}{8\pi G}\int_{\partial\mathcal{M}}d^2x\sqrt{\det(h)}\text{Tr}K+\frac{1}{8\pi G}\int_{\partial\mathcal{M}}d^2x\sqrt{\det(h)},
\end{equation}
where $\partial\mathcal{M}$ is the cutoff surface and $\cal M$ is remaining bulk space after cutoff, $K$ is the extrinsic curvature \footnote{The sign convention for the extrinsic curvature is $K = \frac{1}{2}\mathcal{L}_n g$ where $n$ is a unit normal.} and $h$ is the induced metric on the surface. The variation of the on-shell action with respect to the induced metric gives the Brown-York tensor\footnote{To be more precise, a generalized Brown-York tensor with the contribution of local counter terms included.},
\begin{align}
\delta I_{G-on shell}&= -\frac{1}{2}\int_{\partial\mathcal{M}}d^2x\sqrt{\det(h)}{T^{BY}}^{ij}\delta g_{ij}, \notag\\
    {T^{BY}}^{ij} &= -\frac{1}{8\pi G}(K^{ij} - K g^{ij} + g^{ij}). \label{variation of I}
\end{align}
By the Gauss-Codazzi equation for hypersurfaces in an Einstein space,
\begin{align} \label{Gauss-Codazzi}
    &{\nabla^{(h)}}^i K_{ij} - {\nabla^{(h)}}_j K_i^i = 0, \notag \\
    &(K_i^i)^2-K_{ij}K^{ij} = R(h) + 2,
\end{align}
we have the conservation equation of the Brown-York tensor
\begin{align} \label{BY tensor conservation}
    \nabla^{(h)}_i {T^{BY}}^{ij} = 0,
\end{align}
and the trace relation
\begin{align} \label{BY tensor trace relation}
    {T^{BY}}_i^i = \frac{R(h)}{16\pi G} + 4\pi G ({T^{BY}}^{ij}{T^{BY}}_{ij} - ({T^{BY}}_i^i)^2).
\end{align}

It's customary to work in the Fefferman-Graham coordinates near the conformal boundary, in which the bulk metric takes the form
\begin{align}
    ds^2 = \frac{dy^2}{y^2} + \frac{1}{y^2}g_{ij}(x,y) dx^i dx^j,
\end{align}
and the cutoff surface is taken as $y=y_c$. The background metric of the conformal field theory is
\begin{align}
    \gamma_{ij} = \lim_{y_c\to 0}y_c^2 h_{ij} = g_{ij}|_{y=0}.
\end{align}
In dimension three, the Fefferman-Graham series of the metric truncates as \cite{Skenderis:1999nb}
\begin{equation}
    g_{ij}(x,r)=g^{(0)}_{ij}(x) + y^2 g^{(2)}_{ij}(x)  + y^4 g^{(4)}_{ij}(x) , \label{gij FG expansion}
\end{equation}
The Einstein equation is reduced to one equation that determines $g^{(4)}$ in terms of $g^{(0)}$ and $g^{(2)}$:
\begin{equation}
    g^{(4)}_{ij}=\frac{1}{4}g^{(2)}_{ik}g^{(0)kl}g^{(2)}_{lj},
\end{equation}
and another two equations
\begin{align}
    {\nabla^{(0)}}^i g^{(2)}_{ij} &= \nabla^{(0)}_j {g^{(2)}}_i^i, \label{3D Einstein eqn in FG 1}\\
    {g^{(2)}}_i^i &= -\frac{1}{2}R[g^{(0)}], \label{3D Einstein eqn in FG 2}
\end{align}
where the covariant derivative and raising (lowering) indices are all with respect to the metric $g^{(0)}$. $g^{(0)}$ is identified with the CFT background metric $\gamma$, and the one-point correlator is
\begin{equation}
    \langle T_{ij} \rangle = \lim_{r_c\to 0} {T^{BY}}_{ij}=\frac{1}{8\pi G}\Big(g^{(2)}_{ij}-g^{(0)kl}g^{(2)}_{kl}g^{(0)}_{ij}\Big). \label{BY tensor}
\end{equation}
As in other dimensions, the metric and one-point correlator are sufficient to reconstruct the bulk geometry,
\begin{align} \label{Tgamma to g0g2}
    g^{(0)}_{ij} &= \gamma_{ij}, \notag\\
    g^{(2)}_{ij} &= 8\pi G(\langle T_{ij} \rangle - \langle T_k^k \rangle \gamma_{ij}).
\end{align}
The two equations (\ref{3D Einstein eqn in FG 1}) and (\ref{3D Einstein eqn in FG 2}) are now translated to the conservation of one point function
\begin{align} \label{one-point correlator conservation}
    \nabla^i \langle T_{ij} \rangle = 0,
\end{align}
and the holographic Weyl anomaly
\begin{align} \label{Weyl anomaly}
    \langle T_i^i \rangle = \frac{1}{16\pi G} R[\gamma],
\end{align}
which can also be obtained by taking the $y_c\to 0$ limit of the conservation equation (\ref{BY tensor conservation}) and trace relation (\ref{BY tensor trace relation}) of the Brown-York tensor.

From (\ref{correlator from generating functional}), we can compute multi-point correlators by taking the functional derivative of the one-point correlator. To this end, we give a small variation of the background metric $\gamma$ and solve the variation of one point function order by order from (\ref{one-point correlator conservation}) and (\ref{Weyl anomaly}), which is essentially perturbatively solving the Einstein's equation in the bulk in the Fefferman-Graham coordinates near the conformal boundary. {(In any dimension\footnote{With different Weyl anomaly terms depending on the dimension}, (\ref{one-point correlator conservation}) and (\ref{Weyl anomaly}) represent near boundary analysis and are not sufficient to fully determine the one-point correlator. It needs to be determined from the global geometry.)} In other words, we must require the solution to the Einstein equation, computed in the Fefferman-Graham coordinates near the conformal boundary, to be a global solution. As expected from \cite{Anderson:2004yi}, we get a unique solution of the one-point correlator to the order we compute. Specific to the dimension $d=2$ we are considering, there is an alternative way to calculate the holographic correlators to circumvent the somewhat lengthy geometric computation completely. By relating a change in the modular parameter of the torus to a global variation of the metric \cite{He:2023hoj}, we get a relation between $n+1$ point correlators and $n$ point correlators, which is sufficient to fix the part left undetermined by (\ref{one-point correlator conservation}) and (\ref{Weyl anomaly}) and obtain a recurrence relation of the correlators. The two approaches are elaborated in the following subsections.

\subsection{Holographic two-point correlators in BTZ black hole} \label{2.1}
To compute the holographic correlators, we must consider all classical gravity saddles. To begin with, in this subsection, we evaluate the holographic correlators in the BTZ black hole, which is dominant in the high-temperature limit \cite{Kraus:2006wn}. We will discuss the contribution of other classical saddles in the third subsection. 

The non-rotating BTZ black hole in Euclidean signature is a solution of Einstein's equation \cite{Banados:1992wn, Banados:1992gq},
\begin{equation}
	ds^2=\frac{1}{r^2-r_+^2}dr^2+(r^2-r_+^2)dt^2+r^2d\phi^2, \label{BTZ metric 1}
\end{equation}
where $\phi$ and $t$ are periodic with identifications $\phi\sim \phi+1$ and $t\sim t+1/r_+$.
After rotating to the Lorentzian signature it describes a black hole with an event horizon at $r=r_+$. 

With a coordinate transformation
\begin{align}\label{rho z zbar coordinate}
    \rho=&\ln (r+\sqrt{r^2-r_+^2}),\ \ z=\frac{\phi+it}{2\pi},\ \ \ \ \bar{z}=\frac{\phi-it}{2\pi}.
\end{align}
we obtain the Fefferman-Graham expansion for the Euclidean BTZ black hole metric (\ref{BTZ metric 1}) for $\rho>\ln{r_+}$,
\begin{equation}
	ds^2=d\rho^2+\pi^2 e^{2\rho}\Big[dzd\bar z-e^{-2\rho}(\frac{1}{\tau^2}dz^2+\frac{1}{\bar\tau^2}d\bar z^2)+e^{-4\rho}\frac{1}{\tau^2\bar\tau^2}dzd\bar z\Big].  \label{BTZ metric 2}
\end{equation}
with the identification $z\sim z+1 \sim z+\tau$, where $\tau=i/r_+$. Euclidean BTZ black hole has the topology of a solid torus. The boundary is a two-dimensional torus at $\rho = \infty$. The torus is spanned by two independent cycles, namely, $t$-cycle and $\phi$-cycle. The $t$-cycle is contractible in the bulk while $\phi$-cycle is not. The horizon is the $\phi$-circle at $\rho = \rho_0 = \ln{r_+}$, on which the $t$-cycle contracts into a point. Note that the metric components in $\rho,z,\bar{z}$ coordinates are degenerate at the horizon. This implies that the horizon can not be covered by the coordinate chart $(\rho,z,\bar{z})$ so we need another smoothly compatible coordinate chart
\begin{align}\label{phi x y coordinate}
    \phi &= \phi, \notag\\
    x &= \ln (\frac{r+\sqrt{r^2-r_+^2}}{r_+}) \cos \pi (\frac{z}{\tau}+\frac{\bar{z}}{\bar{\tau}}), \notag\\
    y &= \ln (\frac{r+\sqrt{r^2-r_+^2}}{r_+}) \sin \pi (\frac{z}{\tau}+\frac{\bar{z}}{\bar{\tau}}),
\end{align}
in which the metric is manifestly regular at the horizon.


 We can read off $g^{(0)}$ and $g^{(2)}$ from (\ref{BTZ metric 2}) and obtain the Brown-York tensor of BTZ black hole using (\ref{BY tensor}):
\begin{equation}
	T_{zz}=-\frac{\pi}{8G\tau^2},\ \ T_{\bar z\bar z}=-\frac{\pi}{8G\bar\tau^2},\ \ T_{z\bar z}=T_{\bar zz}=0. \label{BY BTZ 0}
\end{equation}
By the Brown-Henneaux relation \cite{Brown:1986nw} $c=\frac{3}{2G}$ we have
\begin{equation}
	T_{zz}=-\frac{\pi c}{12\tau^2},\ \ T_{\bar z\bar z}=-\frac{\pi c}{12\bar\tau^2},\ \ T_{z\bar z}=T_{\bar zz}=0. \label{BY BTZ}
\end{equation}
The variations of $T^{}_{ij}$ for boundary metric can be obtained by solving the Einstein equation, corresponding to the connected multi-point correlators. Take a variation of the boundary metric,
\begin{align}\label{perturbed boundary metric}
    \gamma_{ij} dx^i dx^j = dz d\bar{z} + \epsilon f_{ij} dx^i dx^j,
\end{align}
where $\epsilon$ is an infinitesimal parameter. We expand the variation of the one-point function in powers of $\epsilon$,
\begin{align}\label{perturbed one-point function}
    \sum_{n=1}^\infty \epsilon^n T^{[n]}_{ij},
\end{align}
plug into (\ref{one-point correlator conservation}) and (\ref{Weyl anomaly}), then perturbatively solve order by order. By (\ref{Tgamma to g0g2}), the solution of the one-point correlator, together with the varied boundary metric, determines a solution to Einstein's equation in the bulk in which $\rho,\phi,t$ or $y,z,\bar{z}$ remains the Fefferman-Graham coordinates of the varied bulk metric. A generic solution of the bulk metric may not take $\rho,\phi,t$ as its Fefferman-Graham coordinates, \footnote{For example, with non-vanishing component $g_{\rho\phi}$} and it is obtained by a boundary-preserving diffeomorphism, which we characterize by a vector expanded in powers of $\epsilon$,
\begin{align}
    V = \sum_{n=1}^\infty \epsilon^n V^{[n]}.
\end{align}
To the first order in $\epsilon$, we find
\begin{align} \label{T1 trace eqn}
    T^{[1]}_{z\bar{z}} = T^{[0]}_{\bar{z}\bar{z}} f_{zz} +2 T^{[0]}_{z\bar{z}} f_{z\bar{z}} + T^{[0]}_{zz} f_{\bar{z}\bar{z}} +\frac{1}{16\pi G}(\partial_{\bar{z}}^2 f_{zz} - 2\partial_z\partial_{\bar{z}} f_{z\bar{z}} + \partial_z^2 f_{\bar{z}\bar{z}}),
\end{align}
and
\begin{align}
    \partial_{\bar{z}}T^{[1]}_{zz} + \partial_z T^{[1]}_{z\bar{z}}  =&  T^{[0]}_{\bar{z}\bar{z}} \partial_z f_{zz} +3 T^{[0]}_{zz} \partial_z f_{\bar{z}\bar{z}} +2 T^{[0]}_{z\bar{z}}( \partial_{\bar{z}} f_{zz} +\partial_z f_{z\bar{z}} ), \nonumber\\
    &+2( f_{zz} \partial_{\bar{z}} T^{[0]}_{z\bar{z}} + f_{z\bar{z}} \partial_{\bar{z}} T^{[0]}_{zz} +f_{z\bar{z}} \partial_z T^{[0]}_{z\bar{z}} +f_{\bar{z}\bar{z}} \partial_z T^{[0]}_{zz})\label{T1 conservation eqn1}\\
    \partial_{z}T^{[1]}_{\bar z\bar z} + \partial_{\bar z} T^{[1]}_{z\bar{z}}=& \text{ c.c. of }(\partial_{\bar{z}}T^{[1]}_{zz} + \partial_z T^{[1]}_{z\bar{z}}). \label{T1 conservation eqn2}
\end{align}
Plugging (\ref{T1 trace eqn}) into (\ref{T1 conservation eqn1}) and using the fact that $\partial_k T^{[0]}_{ij}=0=T^{[0]}_{z\bar{z}}$, we obtain an equation for $T^{[1]}_{zz}$,
\begin{align}
    \partial_{\bar{z}} T^{[1]}_{zz} =2T^{[0]}_{zz} \partial_z f_{\bar{z}\bar{z}} -\frac{1}{16\pi G}( \partial_z\partial_{\bar{z}}^2 f_{zz} - 2\partial_z^2\partial_{\bar{z}} f_{z\bar{z}} + \partial_z^3 f_{\bar{z}\bar{z}}).
\end{align}
We solve this equation by Green's function on the torus 
\begin{align}
    T^{[1]}_{zz}(z) = \frac{1}{16\pi G}
    \{ (-\partial_z\partial_{\bar{z}}f_{zz} + 2 \partial_z^2 f_{z\bar{z}})(z) + \frac{1}{\pi}\int_{\text{T}^2} d^2w G_\tau(z-w) (-\frac{4\pi^2}{\tau^2} \partial_w - \partial_w^3) f_{\bar{z}\bar{z}}(w) + C^{[1]}
    \},
\end{align}
where $C^{[1]}$ is a constant of integration, and the Green's function is
\begin{align}
    G_\tau(z) = \zeta_\tau (z) - 2\zeta_\tau(\frac{1}{2}) z + \frac{2\pi i}{\text{Im}\tau}\text{Im}z,
\end{align}
with $\zeta_\tau (z)$ being the Weierstrass zeta function, see the appendix (\ref{A}) for details. Similarly, we have
\begin{align}
    T^{[1]}_{\bar{z}\bar{z}}(z) = \frac{1}{16\pi G}
    \{ (-\partial_z\partial_{\bar{z}}f_{\bar{z}\bar{z}} + 2 \partial_{\bar{z}}^2 f_{z\bar{z}})(z) + \frac{1}{\pi}\int_{\text{T}^2} d^2w \overline{G_\tau(z-w)} (-\frac{4\pi^2}{\bar\tau^2} \partial_{\bar{w}} - \partial_{\bar{w}}^3) f_{zz}(w) + \bar{C}^{[1]} 
    \}.
\end{align}
By (\ref{Tgamma to g0g2}), we have the first order variation of $g^{(0)}$,
\begin{align} \label{holographic reconstruction g0 first order}
    g^{(0)[1]}_{ij} = f_{ij},
\end{align}
and $g^{(2)}$,
\begin{align} \label{holographic reconstruction g2 first order}
    g^{(2)[1]}_{zz} &= 8\pi G T^{[1]}_{zz}, \notag\\
    g^{(2)[1]}_{z\bar{z}} &= 8\pi G (T^{[0]}_{\bar{z}\bar{z}} f_{zz}+ T^{[0]}_{zz} f_{\bar{z}\bar{z}}), \notag\\
    g^{(2)[1]}_{\bar{z}\bar{z}} &= 8\pi G T^{[1]}_{\bar{z}\bar{z}},
\end{align}
from which we can compute the variation of the bulk metric in its Fefferman-Graham coordinates.

{
Any choice of the constants $C^{[1]}$ and $\bar{C}^{[1]}$ and any vector $V^{[1]}$ defined in the region $\rho\in(\ln r_+,\infty)$ represents, to the first order of $\epsilon$, a solution of the bulk metric manifestly regular in the region $\rho\in(\ln r_+,\infty)$. The condition of being a global solution is equivalent to the regularity at the horizon, which determines the constants of integration.

We have solved the metric with respect to the variation of boundary metric in the new Fefferman-Graham coordinate $(x^{\mu'}) = (\rho',z',\bar{z}')$. To analyze the regularity at the horizon, we should get the components of the varied metric in a coordinate that covers the horizon e.g. $(x^{\tilde{\mu}})=(\phi,x,y)$ and let them be regular at the horizon. For our purpose, it is more convenient to consider the variation of the metric
\begin{align}
    \delta ds^2 =g'_{\mu\nu}(x^{\sigma'})dx^{\mu'} dx^{\nu'} - g_{\mu\nu}(x^\sigma) dx^\mu dx^\nu,
\end{align}
which should be regular as well. We first change $(x^{\mu'})$ into $(x^\mu)$ to the first order of $\epsilon$, such as
\begin{align}
    \delta ds^2 =& g'_{\mu\nu}(x^{\sigma'}) dx^{\mu'} dx^{\nu'} -g_{\mu\nu}(x^{\sigma'})dx^{\mu'} dx^{\nu'} +g_{\mu\nu}(x^{\sigma'})dx^{\mu'} dx^{\nu'} - g_{\mu\nu}(x) dx^\mu dx^\nu \nonumber\\
    =& \epsilon g_{ij}^{FG[1]}dx^i dx^j -\epsilon \mathcal{L}_{V^{[1]}} ds^2 +O(\epsilon^2),
\end{align}
and then change $(x^\mu)$ into $(x^{\tilde{\mu}})$ in which the components of the metric at the horizon should be regular, i.e.
\begin{align}\label{global regularity condition}
    g_{ij}^{FG[1]}\frac{\partial x^i}{\partial x^{\tilde{\mu}}} \frac{\partial x^j}{\partial x^{\tilde{\nu}}} -(\mathcal{L}_{V^{[1]}} ds^2)_{\tilde{\mu}\tilde{\nu}}
\end{align}
should be regular at $(\phi,x=0,y=0)$. $g_{ij}^{FG[1]}$ comes from the solution in the F-G gauge with the boundary variation, which is exactly what we solved above. The second term is the Lie derivative with respect to the diffeomorphism from the old F-G coordinate to the new F-G coordinate.
}

Leaving the detailed computation to the appendix (\ref{B}), we find the necessary condition for the global regularity condition is
\begin{align} \label{equivalent form of the global regularity condition}
    &\int_{\text{T}^2} d^2z (\tau^2 g^{FG[1]}_{zz} - \bar{\tau}^2 g^{FG[1]}_{\bar{z}\bar{z}})\big|_{\rho=\rho_0} = 0 \notag\\
    &\int_{\text{T}^2} d^2z \partial_{\rho}^2(\tau^2 g^{FG[1]}_{zz} + 2\tau \bar{\tau} g^{FG[1]}_{z\bar{z}} + \bar{\tau}^2 g^{FG[1]}_{\bar{z}\bar{z}})\big|_{\rho=\rho_0} = 0,
\end{align}
which determines the constants
\begin{align}
    C^{[1]} &= \frac{4\pi^2 \bar{\tau}}{\tau^3 \text{Im}\tau } \int_{\text{T}^2} d^2z f_{\bar{z}\bar{z}}, \notag\\
    \bar{C}^{[1]} &= \frac{4\pi^2 \tau}{\bar{\tau}^3 \text{Im}\tau} \int_{\text{T}^2} d^2z f_{zz}.
\end{align}
Now the first-order variation of the one-point correlator is completely determined. By the equation
\begin{equation}
	\langle{T_{ij}(z_1)T_{kl}(z_2)}\rangle=-\frac{2}{\sqrt{\det(\gamma(z_2))}}\frac{\delta \langle T_{ij}(z_1) \rangle}{\delta\gamma^{kl}(z_2)} + \frac{\gamma_{kl}(z_1)}{\sqrt{\det(\gamma(z_1))}}\langle T_{ij}(z_1) \rangle \delta(z_1-z_2) ,\label{2pt general}
\end{equation}
we obtain all the holographic two-point correlators of the stress tensor.

The independent two-point correlators in the BTZ black hole are listed below: 
\begin{align}
	\langle{T(z_1)T(z_2)}\rangle=&2\pi i\partial_{\tau}\langle{T}\rangle+\frac{c}{12}\Big[\frac{4\pi^2}{\tau^2}(\wp(z_1-z_2)+2\eta)+\wp''(z_1-z_2)\Big], \label{BTZ TT}\\
	\langle{T(z_1)\bar T(z_2)}\rangle=&2\pi i\partial_{\tau}\langle{\bar T}\rangle,\label{BTZ TTbar} \\
	\langle{\Theta(z_1)T(z_2)}\rangle=&0,\label{BTZ Ttheta} \\
	\langle{\Theta(z_1)\Theta(z_2)}\rangle=&0,\label{BTZ thetatheta} 
\end{align}
where $\wp_\tau(z) = -\zeta_\tau^{'}(z)$ is the Weierstrass $P$ function and we used the convention  $T=-2\pi T_{zz},\, \bar T=-2\pi T_{\bar z\bar z},\, \Theta=-2\pi T_{z\bar z}$, the central charge relation $c=\frac{3}{2G}$ and the explicit expression of the Green's function. 

\subsection{Recurrence relation of correlators} \label{2.2}
In principle, with increasingly tedious computation, we can compute higher order variation of the one-point correlator to obtain higher point correlators as described in the previous section. As discussed in the introduction, the alternative way of computing the correlators is to derive a recurrence relation following the spirit of \cite{Friedan:1986ua, Eguchi:1986sb}.

We set the metric of the CFT to be in Polyakov's light cone gauge,
\begin{align}\label{Polyakov's light cone gauge}
    ds^2 = dzd\bar{z} + F(z) d\bar{z}^2,
\end{align}
that is, we turn on a variation of $\gamma_{\bar{z}\bar{z}}$ of the Euclidean metric while keeping other components fixed. Then (\ref{Weyl anomaly}) is reduced to
\begin{align} \label{Weyl anomaly for Virasoro Ward identity}
    \langle T_{z\bar{z}} \rangle = F \langle T_{zz} \rangle + \frac{1}{16\pi G} \partial_z^2 F,
\end{align}
and the first equation of (\ref{one-point correlator conservation}) is reduced to
\begin{align} \label{one-point correlator conservation for Virasoro Ward identity}
    \partial_{\bar{z}} \langle T_{zz} \rangle - 3\partial_z F \langle T_{zz} \rangle -2 F \partial_z \langle T_{zz} \rangle + \partial_z \langle T_{z\bar{z}} \rangle= 0.
\end{align}
Plugging (\ref{Weyl anomaly for Virasoro Ward identity}) into (\ref{one-point correlator conservation for Virasoro Ward identity}), we obtain the holographic Virasoro Ward identity \cite{Polyakov:1987zb,Banados:2004nr}
\begin{align}
    \partial_{\bar{z}} \langle T_{zz} \rangle - 2 \partial_z F \langle T_{zz} \rangle - F \partial_z \langle T_{zz} \rangle + \frac{1}{16\pi G} \partial_z^3 F = 0.
\end{align}
Taking $n$-th functional derivatives of $F$ and evaluate at $F=0$, we find that for $n=1$,
\begin{align}
    \partial_{\bar{z}}\langle T_{zz}(z)T_{zz}(z_1)\rangle - \partial_z\delta(z-z_1)\langle T_{zz}(z) \rangle - \frac{1}{2} \delta(z-z_1)\partial_z\langle T_{zz}(z) \rangle + \frac{1}{32\pi G} \partial_z^3 \delta(z-z_1) = 0,
\end{align}
and for $n \geq 2$,
\begin{align}
    &\partial_{\bar{z}} \langle T_{zz}(z) T_{zz}(z_1) \ldots T_{zz}(z_n) \rangle - \sum_{i=1}^n \partial_{z}\delta(z-z_i) \langle T_{zz}(z) T_{zz}(z_1) \ldots T_{zz}(z_{i-1})T_{zz}(z_{i+1}) \ldots T_{zz}(z_n) \rangle \nonumber\\
    &- \frac{1}{2} \sum_{i=1}^n \delta(z-z_i) \partial_z \langle T_{zz}(z) T_{zz}(z_1) \ldots T_{zz}(z_{i-1})T_{zz}(z_{i+1}) \ldots T_{zz}(z_n) \rangle = 0.
\end{align}
Solving the equation with the Green's function on a torus, we find that for $n=1$,
\begin{align}\label{2-pt correlator with unspecified constant}
    &\langle T_{zz}(z)T_{zz}(z_1)\rangle = \frac{1}{\pi} [\partial_z G(z-z_1) \langle T_{zz}(z_1) \rangle - \frac{1}{2} G(z-z_1) \partial_{z_1}\langle T_{zz}(z_1) \rangle - \frac{1}{32\pi G}\partial_z^3 G(z-z_1)] \notag\\
    &+\frac{1}{\text{Im}\tau}\int_{\text{T}^2} d^2v \langle T_{zz}(v)T_{zz}(z_1)\rangle,
\end{align}
and for $n \geq 2$,
\begin{align}\label{n-pt correlator with unspecified constant}
    &\langle T_{zz}(z) T_{zz}(z_1) \ldots T_{zz}(z_n) \rangle =  \frac{1}{\pi} \sum_{i=1}^n [\partial_z G(z-z_i) \langle T_{zz}(z_1) \ldots T_{zz}(z_n) \rangle \notag\\
    &- \frac{1}{2} G(z-z_i) \partial_{z_i} \langle T_{zz}(z_1) \ldots T_{zz}(z_n) \rangle] + \frac{1}{\text{Im}\tau} \int_{\text{T}^2} d^2v \langle T_{zz}(v) T_{zz}(z_1) \ldots T_{zz}(z_n) \rangle.
\end{align}
The last terms on the right-hand sides of (\ref{2-pt correlator with unspecified constant}) and (\ref{n-pt correlator with unspecified constant}) are one-point-averaged correlators and can be computed via the global metric variation of fewer-point-correlators,
\footnote{
`global' means that the metric variation does not depend on the coordinate, i.e. $\delta\gamma_{ij}$ is constant.
}
while the other terms depend only on the near-boundary analysis and fewer point correlators. 
And in the following, we will show that the regularity condition for the global metric variation can be further transformed to that for a modular variation.

To this end, we exploit the fact that a variation of the modular parameter is equivalent to the combined operation of a global metric variation of $\gamma_{zz}$ and $\gamma_{\bar{z}\bar{z}}$, a coordinate transformation (diffeomorphism), and a global Weyl transformation. To be precise, let the global metric variation be $\delta\gamma_{\bar z\bar z}(z)=\alpha, \, \delta\gamma_{zz}(z)=\bar\alpha$, where $\alpha$ is an infinitesimal complex constant. To the first order in $\alpha$, the new metric is
\begin{align}
	ds^2=&dzd\bar z+\bar\alpha dz^2+\alpha d\bar z^2\notag\\
	=&(1+\alpha+\bar\alpha)d(z+\alpha(\bar z-z))d(\bar z+\bar\alpha(z-\bar z)).
\end{align}
With a Weyl transformation of factor $(1-\alpha-\bar{\alpha})$ and a change of coordinates
\begin{equation}
	z'=z+\alpha(\bar z-z),\ \ \bar z'=\bar z+\bar\alpha(z-\bar z),\label{coordinate transformation}
\end{equation}
we get a torus with the flat metric $ds^2= dz'd\bar{z}'$ and a varied modular parameter
\begin{align}\label{modular variation}
    \tau' = \tau + \alpha (\bar{\tau} -\tau).
\end{align}
According to \cite{Banados:1998gg}, there exists a bulk coordinate chart $(\rho',z',\bar{z}')$ in which the bulk metric with such boundary takes the Banados form 
\begin{equation}
	ds^2=d\rho'^2+\pi^2 e^{2\rho'}\Big[dz'd\bar z'-e^{-2\rho'}(\mathcal{L} dz'^2 +\bar{\mathcal{L}} d\bar z'^2)+e^{-4\rho'}\mathcal{L} \bar{\mathcal{L}} dz'd\bar z'\Big].\label{Banados form}
\end{equation}
Due to the translation symmetry $(\rho',z',\bar{z}')\to(\rho',z'+\lambda,\bar{z}'+\bar{\lambda})$, $\mathcal{L}$ and $\bar{\mathcal{L}}$ are constants. However, for the general value of $\mathcal{L}$ and $\bar{\mathcal{L}}$, this metric may cause conical singularity at the the circle $\rho'=\frac{1}{4}\ln(\mathcal{L} \bar{\mathcal{L}})$. 

To avoid this conical singularity, the modular parameters $\tau',\bar{\tau}'$ of the boundary torus should have a relation with $\mathcal{L}$ and $\bar{\mathcal{L}}$. To see this, let us expand the bulk metric around $\rho'=\frac{1}{4}\ln(\mathcal{L} \bar{\mathcal{L}})$,
\begin{align}
	ds^2=&d\rho'^2 - \pi^2 (\sqrt{\mathcal{L}} dz'-\sqrt{\bar{\mathcal{L}}} d\bar{z}')^2 + 4 \pi^2 \sqrt{\mathcal{L} \bar{\mathcal{L}}} (\rho-\frac{\ln (\mathcal{L} \bar{\mathcal{L}})}{4})^2 dz' d\bar{z}' +O(\rho - \frac{\ln (\mathcal{L} \bar{\mathcal{L}})}{4})^3.
\end{align}
Then we take a coordinate transformation
\begin{align}
    \tilde{\phi} = (\sqrt{\mathcal{L}} z'-\sqrt{\bar{\mathcal{L}}} \bar{z}')/i, \quad
    \tilde{t} = \sqrt{\mathcal{L}} z'+\sqrt{\bar{\mathcal{L}}} \bar{z}',
\end{align}
consider a slice at $\tilde{\phi}=const.$, and the induced metric on this slice is 
\begin{align}
	ds^2=&d\rho'^2 + (\rho'-\frac{\ln (\mathcal{L} \bar{\mathcal{L}})}{4})^2 d(\pi \tilde{t})^2 +O(\rho' - \frac{\ln (\mathcal{L} \bar{\mathcal{L}})}{4})^3, 
\end{align}
The absence of the conical singularity requires a period on the slice that $(\rho',\tilde{\phi},\pi\tilde{t}) \sim (\rho',\tilde{\phi},\pi\tilde{t}+2\pi) $. On the other hand, all the periods for the bulk solid torus can be expressed in $(\rho',z',\bar{z}')$ coordinate as $(\rho',z,\bar{z})\sim (\rho',z'+m+n\tau',\bar{z}'+m+n\bar{\tau}')$, for arbitrary integers $m$ and $n$, therefore $(\rho',\tilde{\phi},\tilde{t}) \sim (\rho',\tilde{\phi},\tilde{t}+2)$ must correspond to some of them. Different correspondences by choosing various $m$ and $n$ provide different spacetime saddles that are related by modular transformations. For example, having the period $(\tilde{\phi},\tilde{t}) \sim (\tilde{\phi},\tilde{t}+2) $ correspond with $z'\sim z'+\tau'$ makes the geometry a BTZ black hole. 
\footnote{The thermal $\rm AdS$ saddle corresponds to the case of $m=1$, $n=0$, i.e. $z'\sim z'+1$.}
More precisely, such correspondence gives us two equation for $\sqrt{\mathcal{L}}$ and $\sqrt{\bar{\mathcal{L}}}$,
\begin{align}
    \sqrt{\mathcal{L}} \tau'-\sqrt{\bar{\mathcal{L}}} \bar{\tau}'=0, \quad
    \sqrt{\mathcal{L}} \tau'+\sqrt{\bar{\mathcal{L}}} \bar{\tau}'=2,
\end{align}
The solution is 
\begin{equation}
    \mathcal{L}=\frac{1}{\tau'^2}, \quad
    \bar{\mathcal{L}}=\frac{1}{\bar{\tau}'^2},
\end{equation}
and the according bulk metric is 
\begin{equation}\label{bulk metric after modular variation}
	ds^2=d\rho'^2+\pi^2 e^{2\rho'}\Big[dz'd\bar z'-e^{-2\rho'}(\frac{1}{\tau'^2} dz'^2 +\frac{1}{\bar{\tau}'^2} d\bar z'^2)+e^{-4\rho'}\frac{1}{\tau'^2 \bar{\tau}'^2} dz'd\bar z'\Big].
\end{equation}

Since an infinitesimal perturbation on the boundary should not lead to a finite change in the bulk, the bulk stays, after the boundary modular variation (\ref{modular variation}), in the BTZ black hole saddle, and accordingly, the metric becomes (\ref{bulk metric after modular variation}). That is to say, when the boundary varies infinitesimally as (\ref{modular variation}), to keep the geometry globally regular, the bulk metric should change correspondingly 
\begin{align}\label{dependence of g(2) on varied tau}
    g^{(2)}_{ij}\to g^{(2)}_{ij}{}'= -\pi^2 l^2
    \begin{pmatrix}
        \frac{1}{\tau'{}^2} & 0 \\
        0 & \frac{1}{\bar{\tau}'{}^2}
    \end{pmatrix},
\end{align} 
which amounts to
replace $\tau$ with $\tau'$ in the Euclidean BTZ black hole metric (\ref{BTZ metric 2}). 

Since a diffeomorphism always reserves regularity, the global metric variation behaves regularly if and only if the metric is regular after the change of modular parameter. So we can transform the regularity condition of the varied metric into the regularity condition under a change of the modular parameter. More precisely, let $O$ be an arbitrary operator and then its change under a variation of the metric and of the modular parameter has a relation that
\begin{align}\label{equivalent relation}
    \int_{\rm T^2} d^2 z \delta \gamma_{ij} \frac{\delta \langle O \rangle}{\delta \gamma_{ij}} + \alpha \mathcal{L}_{\xi}\langle O \rangle + \bar{\alpha} \mathcal{L}_{\bar{\xi}}\langle O \rangle = \delta \tau \partial_\tau \langle O \rangle + \delta \bar{\tau} \partial_{\bar\tau} \langle O \rangle,
\end{align}
with $\delta \gamma_{ij}$, $\xi$, $\bar{\xi}$, $\delta \tau$ and $\delta \bar{\tau}$ taking the following forms,
\begin{gather}
    \delta \gamma_{zz} = \bar{\alpha}, \quad \delta \gamma_{\bar{z}\bar{z}} = \alpha, \quad \delta \gamma_{z\bar{z}} = - \alpha - \bar{\alpha}, \\
    \xi = (z-\bar{z})\partial_z, \quad \bar{\xi} =(\bar{z}-z)\partial_{\bar{z}}, \\
    \delta \tau = \alpha (\bar{\tau} - \tau), \quad \delta \bar{\tau} = \bar{\alpha} (\tau - \bar{\tau}).
\end{gather}
And it is also worth noting that the dependence of $\langle O \rangle$ on the boundary modular parameters $\tau$ and $\bar{\tau}$ comes from the dependence of $g^{(2)}_{ij}$ on them, i.e. (\ref{dependence of g(2) on varied tau}).

Since $\alpha$ and $\bar{\alpha}$ are independent, (\ref{equivalent relation}) provides two independent equations respectively,
\begin{align}
    (\bar{\tau}-\tau) \partial_\tau \langle O \rangle &= \mathcal{L}_{(z-\bar{z})\partial_z} \langle O \rangle
    + \int_{\text{T}^2} d^2z \big( \frac{\delta \langle O \rangle}{\delta \gamma_{\bar{z}\bar{z}}(z)} -  \frac{\delta \langle O \rangle}{\delta \gamma_{z\bar{z}}(z)} \big) \label{global metric variation and modular differentiation eqn 1},\\
    (\tau-\bar{\tau}) \partial_{\bar{\tau}} \langle O \rangle &= \mathcal{L}_{(\bar{z}-z)\partial_{\bar{z}}} \langle O \rangle
    + \int_{\text{T}^2} d^2z \big( \frac{\delta \langle O \rangle}{\delta \gamma_{zz}(z)} -  \frac{\delta \langle O \rangle}{\delta \gamma_{z\bar{z}}(z)} \big). \label{global metric variation and modular differentiation eqn 2}
\end{align}

In principle, the term corresponds to the global Weyl transformation
\begin{align}
    \int_{\text{T}^2} d^2z \frac{\delta \langle T_{i_1j_1}(z_1) \ldots T_{i_nj_n}(z_n) \rangle}{\delta \gamma_{z\bar{z}}(z)}
\end{align}
can be expressed in terms of the variation with respect to $\gamma_{zz}$ and $\gamma_{\bar{z}\bar{z}}$ with the help of the holographic Weyl anomaly (\ref{Weyl anomaly}).
However, in the case of CFT, it vanishes because a global Weyl transformation does not change the correlators. In the holographic language, this is true because a global Weyl transformation of $g^{(0)}$ does not change $g^{(2)}$. 

Acting relations (\ref{global metric variation and modular differentiation eqn 1}) and (\ref{global metric variation and modular differentiation eqn 2}) upon $\langle O \rangle = \langle T_{i_1j_1}(z_1) \ldots T_{i_nj_n}(z_n) \rangle$, we obtain two equations on $\int_{\text{T}^2} d^2z \frac{\delta \langle T_{i_1j_1}(z_1) \ldots T_{i_nj_n}(z_n) \rangle}{\delta \gamma_{zz}(z)}$ and $\int_{\text{T}^2} d^2z \frac{\delta \langle T_{i_1j_1}(z_1) \ldots T_{i_nj_n}(z_n) \rangle}{\delta \gamma_{\bar{z}\bar{z}}(z)}$ and then all one point averaged $(n+1)$-point correlators can be computed if we know all $n$-point correlators. 
For example, setting the correlator $\langle O \rangle$ in (\ref{global metric variation and modular differentiation eqn 1}) and (\ref{global metric variation and modular differentiation eqn 2}) to be $\langle T_{zz}(z) \rangle$ we obtain
\begin{align}
    (\bar{\tau}-\tau) \partial_\tau \langle T_{zz}(z) \rangle &= 2 \langle T_{zz}(z) \rangle
    + \int_{\text{T}^2} d^2w \frac{\delta \langle T_{zz}(z) \rangle}{\delta \gamma_{\bar{z}\bar{z}}(w)} ,\\
    (\tau-\bar{\tau}) \partial_{\bar{\tau}} \langle T_{zz}(z) \rangle &= -2 \langle T_{z\bar{z}}(z) \rangle
    + \int_{\text{T}^2} d^2w  \frac{\delta \langle T_{zz}(z) \rangle}{\delta \gamma_{zz}(w)}. 
\end{align}
Plugging in one-point correlators (\ref{BY BTZ 0}), we get
\begin{align}
    \int_{\text{T}^2} d^2w \frac{\delta \langle T_{zz}(z) \rangle}{\delta \gamma_{\bar{z}\bar{z}}(w)} &= \frac{\pi\bar\tau}{4G\tau^3}, \\
    \int_{\text{T}^2} d^2w \frac{\delta \langle T_{zz}(z) \rangle}{\delta \gamma_{zz}(w)} &= 0.
\end{align}
Then we have
\begin{align}
    \int_{\text{T}^2} d^2w \langle T_{zz}(z) T_{zz}(w)\rangle = \frac{\pi\bar\tau}{8G\tau^3}.
\end{align}
Plugging into (\ref{2-pt correlator with unspecified constant}), we recover the result in (\ref{BTZ TT}).

From (\ref{global metric variation and modular differentiation eqn 1}), we find the one-point-averaged correlators are given by \footnote{On the right-hand side, the first and second terms are not periodic in $z_i$ separately, but the combination is. In fact by taking $\tau$ derivative of $\langle T_{zz}(z_1) \ldots T_{zz}(z_k + \tau) \ldots T_{zz}(z_n) \rangle = \langle T_{zz}(z_1) \ldots T_{zz}(z_k) \ldots T_{zz}(z_n) \rangle$, we find
\begin{align}
     &\partial_\tau \langle T_{zz}(z_1) \ldots T_{zz}(z_k^{'}) \ldots T_{zz}(z_n) \rangle|_{z_k^{'}=z_k+\tau} - \partial_\tau \langle T_{zz}(z_1) \ldots T_{zz}(z_k) \ldots T_{zz}(z_n) \rangle \notag\\
     &= -\partial_{z_k}  \langle T_{zz}(z_1) \ldots T_{zz}(z_k) \ldots T_{zz}(z_n) \rangle \notag
\end{align}
As a result, the expression of correlators in the recurrence relation is not term-by-term doubly periodic, but it is as a whole.}
\begin{align}
    &2\int_{\text{T}^2} d^2z \langle T_{zz}(z) T_{zz}(z_1) \ldots T_{zz}(z_n) \rangle = (\bar{\tau} - \tau) \partial_\tau \langle T_{zz}(z_1) \ldots T_{zz}(z_n) \rangle \notag\\
    &+ \sum_{i=1}^n (\bar{z}_i - z_i) \partial_{z_i} \langle T_{zz}(z_1) \ldots T_{zz}(z_n) \rangle - 2n \langle T_{zz}(z_1) \ldots T_{zz}(z_n) \rangle.\label{65}
\end{align}
Then we obtain the full form of the recurrence relation (\ref{n-pt correlator with unspecified constant}),
\begin{multline}
    \langle T_{zz}(z) T_{zz}(z_1) \ldots T_{zz}(z_n) \rangle = -i \partial_\tau \langle T_{zz}(z_1) \ldots T_{zz}(z_n) \rangle \\
    - \frac{1}{2\pi} \sum_{i=1}^n [2(\wp_\tau(z-z_i)+2\zeta_\tau(\frac{1}{2})) \langle T_{zz}(z_1) \ldots T_{zz}(z_n) \rangle \\
    + (\zeta_\tau(z-z_i) - 2\zeta_\tau(\frac{1}{2})(z-z_i))\partial_{z_i}\langle T_{zz}(z_1) \ldots T_{zz}(z_n) \rangle], \label{recurrence relation for general saddle}
\end{multline}
or equivalently,
\begin{align}
    \langle T(z) T(z_1) \ldots T(z_n) \rangle = &2\pi i \partial_\tau \langle T(z_1) \ldots T(z_n) \rangle \notag\\
    &+ \sum_{i=1}^n [2(\wp_\tau(z-z_i)+2\zeta_\tau(\frac{1}{2})) \langle T(z_1) \ldots T(z_n) \rangle \notag\\
    &+ (\zeta_\tau(z-z_i) - 2\zeta_\tau(\frac{1}{2})(z-z_i))\partial_{z_i}\langle T(z_1) \ldots T(z_n) \rangle] .
\end{align}
In addition, the one-point correlators conforms to the recurrence relation in its form as well,
\begin{align}
    \langle T_{zz} \rangle &= i \partial_\tau I, \notag\\
    \langle T_{\bar{z}\bar{z}} \rangle &= i \partial_{\bar{\tau}} I, \notag\\
    \langle T_{z\bar{z}} \rangle &= 0.
\end{align}
With the recurrence relation, we can compute correlators of $T_{zz}$ of any points. Then we can use the Ward identity of conservation repeatedly to compute correlators including other components of the stress tensor, most of them being contact terms except for those correlators of all $T$s or all $\bar{T}$s. The resulting expressions are exactly consistent with the pure field theory derivation provided by \cite{He:2020udl}.
In addition, the above recurrence relation (\ref{n-pt correlator with unspecified constant}) is equivalent to the Zhu relation as presented in the vertex operator algebra formalism~\cite{Zhu1996}, which has been rephrased in more physical terms in~\cite{Maloney:2018hdg}.

\subsection{Other classical gravity saddles}\label{2.3}
There are other classical gravity saddles with the torus as its conformal boundary, which are classified in \cite{Maloney:2007ud}. The real smooth geometries, first considered in \cite{Maldacena:1998bw}, are obtained from the thermal AdS by modular transformations
\begin{align}
    \tau \to \frac{a\tau + b}{c\tau + d}
\end{align}
They are labeled by the relative prime pair $(c,d)$ and denoted as $M_{c,d}$ in \cite{Maloney:2007ud}. In particular, the thermal AdS is $M_{1,0}$ and the BTZ black hole is $M_{0,1}$. The derivation of the recurrence relation does not assume the global geometry of the gravity saddle, so the recurrence relation can be used to compute the correlators for any gravity saddle when it dominates in the path integral, for example, the thermal AdS in the low-temperature limit or the BTZ black hole in the high-temperature limit. One can also compute the exact correlators from a full partition function if it's available, though this question is much more subtle and difficult, see \cite{Maloney:2007ud} for a discussion of the full partition function for a minimal gravity theory, and \cite{Yin:2007gv} including higher genus cases.
\section{Correlators obtained from Chern-Simons formalism}
In subsection \ref{2.1}, we calculate the non-trivial two-point correlators $\langle{T(z_1)T(z_2)}\rangle$ by taking variations of Brown-York tensor. To evaluate the correlators with more stress tensors inserted, we provide a recurrence relation between $(n+1)$-point and $n$-point correlators in subsection \ref{2.2}. An alternative approach to derive such a recurrence relation is through utilizing the equation of motion of the stress tensor in Chern-Simons formulation. For this calculation, we mainly refer to the work \cite{Bagchi:2015wna}, in which the recursion relation for the holographic correlators in GCFT on the cylinder has been obtained. \par
Let us briefly review Chern-Simons theory and its relationship to 3D gravity. Chern-Simons theory is a non-Abelian gauge field theory with action
\begin{equation}
	I_{CS}[A]=\frac{k}{4\pi}\int\text{Tr}(A\wedge dA+\frac{2}{3}A\wedge A\wedge A),
\end{equation}
and the variation of the action with respect to $A$ gives the equation of motion
\begin{equation}
	dA+A\wedge A=0. \label{EOM A}
\end{equation}
It has been proved that 3D gravity with a negative cosmological constant is equivalent to Chern-Simons theory with gauge group $SL(2,\mathbb{R}) \times SL(2,\mathbb{R})$, where the Einstein-Hilbert action can be written as the difference between two Chern-Simons actions \cite{Achucarro:1986uwr, Witten:1988hc}
\begin{equation}
	I_{EH}=I_{CS}[A]-I_{CS}[\bar A], \label{CS gravitational action}
\end{equation}
where $k=1/4G=c/6$. The gauge fields $A$ and $\bar A$ are constructed by the frame $e^a_\mu$, the spin connection $\omega^a_\mu$ and the generators $J_{a}$ of the fundamental representation of $sl(2,\mathbb{R})$:
\begin{align}
	A=&(\omega^a_\mu+e^a_\mu)dx^\mu J_{a},\notag\\
	\bar A=&(\omega^a_\mu-e^a_\mu)dx^\mu J_{a},
\end{align}
The metric of 3D gravity takes the form
\begin{equation}
	ds^2=\frac{1}{2}\text{Tr}[(A-\bar A)^2]. \label{3D gravity metric}
\end{equation}
For the asymptotically AdS spacetimes, $A$ and $\bar A$ can be written in the highest-weight gauge \cite{Banados:1998gg,Banados:2004nr} and take the forms $A=b(d+a)b^{-1}$ and $\bar A=b^{-1}(d+\bar a)b$, where $b=e^{\rho L_0}$ for $\rho$ being the radial coordinate. The new gauge fields $a(z,\bar z)$ and $\bar a(z,\bar z)$ are independent of the radial coordinate and have expansions on the $sl(2,\mathbb{R})$ generators \cite{Banados:2004nr},
\begin{align}
	a_z(z,\bar z)=&L_++\frac{\mathcal{T}(z,\bar z)}{k}L_{-},\notag\\
	a_{\bar z}(z,\bar z)=&-\mu(z,\bar z)L_++\alpha L_0+\beta L_{-},\notag\\
	\bar a_{z}(z,\bar z)=&\bar\beta L_{+}+\bar\alpha L_0-\bar\mu(z,\bar z)L_{-},\notag\\
	\bar a_{\bar z}(z,\bar z)=&\frac{\bar{\mathcal{T}}(z,\bar z)}{k}L_{+}+L_{-}, \label{highest weight gauge}
	\end{align}
where the generators take the forms
\begin{equation}
	L_+=
    \begin{bmatrix}
    0&1\\0&0
    \end{bmatrix}
    ,\quad
	 L_0=
    \begin{bmatrix}
    \frac{1}{2}&0\\0&-\frac{1}{2}
    \end{bmatrix}
    ,\quad
    L_{-}=
    \begin{bmatrix}
    0&0\\-1&0
    \end{bmatrix}.
\end{equation}
Plugging the highest-weight gauge into the equation of motion (\ref{EOM A}) implies $\partial_{z}a_{\bar z}-\partial_{\bar z}a_z+[a_z,a_{\bar z}]=0$, which determines $\alpha$, $\beta$ and gives us the equation of motion for $\mathcal{T}$,
\begin{align}
	\alpha=-\partial_z\mu,\ \ \beta=-\mu\frac{\mathcal{T}}{k}-\frac{1}{2}\partial^2_z\mu,\ \ -\partial_{\bar z}\mathcal{T}=\mu\partial_z\mathcal{T}+2\partial_z\mu\mathcal{T}+\frac{k}{2}\partial^3_{z}\mu. \label{EOM T}
\end{align}
One can derive similar equations for $\bar\alpha$, $\bar\beta$ and $\bar{\mathcal{T}}$,
\begin{equation}
	\bar\alpha=\partial_{\bar z}\bar\mu,\ \ 
	\bar\beta=-\bar\mu\frac{\bar{\mathcal{T}}}{k}-\frac{1}{2}\partial^2_{\bar z}\bar\mu,\ \ -\partial_{z}\bar{\mathcal{T}}=\bar\mu\partial_{\bar z}\bar{\mathcal{T}}+2\partial_{\bar z}\bar\mu\bar{\mathcal{T}}+\frac{k}{2}\partial^3_{\bar z}\bar\mu.
\end{equation}
 The boundary metric is obtained by applying (\ref{highest weight gauge}) to (\ref{3D gravity metric}) and keeping the leading terms in the limit $\rho\rightarrow\infty$,
\begin{align}
	ds^2=e^{2\rho}\Big[(1+\mu\bar\mu)dzd\bar z-\mu d\bar z^2-\bar\mu dz^2\Big]+O(1),
\end{align}
where $\mu$ and $\bar\mu$ can be interpreted as boundary metric components $-g_{\bar z\bar z}$ and $-g_{zz}$. The variations of the generating functional\footnote{In fact, the generating functional $I[\mu,\bar\mu]$ here consists of the Einstein-Hilbert term (\ref{CS gravitational action}) and a boundary term:
\begin{equation}
	I[\mu,\bar\mu]=I_{CS}[A]-I_{CS}[\bar A]+\frac{k}{4\pi}\int_{\partial\mathcal{M}}\Big(R+\frac{1}{2}K\Big).
\end{equation}
For a detailed discussion, we refer to \cite{Banados:2004nr}.
} with respect to $\mu$ and $\bar\mu$ give $\mathcal{T}$ and $\bar{\mathcal{T}}$, respectively,
\begin{equation}
	\mathcal{T}=-2\pi\frac{\delta I[\mu,\bar\mu]}{\delta\mu},\ \ \bar{\mathcal{T}}=-2\pi\frac{\delta I[\mu,\bar\mu]}{\delta\bar\mu}.
\end{equation}
Thus $\mathcal{T}$ and $\bar{\mathcal{T}}$ can be interpreted as components $-2\pi T^{\bar z\bar z}$ and $-2\pi T^{zz}$ of Brown-York tensor, which correspond to the one-point correlators $\langle{T}\rangle_{\mu,\bar\mu}$ and $\langle{\bar T}\rangle_{\mu,\bar\mu}$ on the CFT side. The subscript here implies that this boundary CFT has the metric $ds^2=(1+\mu\bar\mu)dzd\bar z-\mu d\bar z^2-\bar\mu dz^2$. In this paper, we focus on the connected $n$-point correlator $\langle{T(z_1)T(z_2)...T(z_n)}\rangle_0$ in CFT with flat metric, which can be read off from the deformed one-point correlator $\langle{T(z_1)}\rangle_\mu$ with the source $\mu(z,\bar z)$ localized at $z_2,z_3,...,z_n$ \cite{Bagchi:2015wna},
\begin{equation}
	\mu(z,\bar z)=\sum_{i=2}^n\epsilon_i\delta(z-z_i),\ \ \bar\mu(z,\bar z)=0, 
\end{equation}
where $\epsilon_i$ is a small parameter. $\mu(z,\bar z)$ can be interpreted as the variation $-\delta g_{\bar z\bar z}$ of the flat metric $ds^2=dzd\bar z$, and thus the deformed CFT action $S_\mu=S_0-\int d^2z\mu(z,\bar z)T(z)=S_0-\sum_{i}\epsilon_iT(z_i)$. The deformed one-point correlator $\langle{T(z_1)}\rangle_\mu$ is computed using the path integral formula
\begin{align}
	\langle{T(z_1)}\rangle_\mu=&\frac{1}{Z_{\mu}}\int D\phi\ T(z_1)e^{-S_\mu[\phi]}\notag\\
	=&\Big(\int D\phi\ \prod_{i=2}^ne^{\epsilon_iT(z_i)}e^{-S_0[\phi]}\Big)^{-1}\int D\phi\ T(z_1)\prod_{i=2}^ne^{\epsilon_iT(z_i)}e^{-S_0[\phi]}\notag\\
	=&\langle{T(z_1)}\rangle_0+\sum_{i=2}^n\epsilon_i\langle{T(z_1)T(z_i)}\rangle_0+\sum_{i,j=2}^n\frac{1}{2}\epsilon_i\epsilon_j\langle{T(z_1)T(z_i)T(z_j)}\rangle_0\notag\\
	&+...+\sum_{i_2,...,i_n=2}^n\frac{1}{(n-1)!}\epsilon_{i_2}...\epsilon_{i_n}\langle{T(z_1)T(z_{i_2})...T(z_{i_n})}\rangle_0+O(\epsilon^n). \label{deformed 1-pt expansion}
\end{align}
The connected $n$-point correlator $\langle{T(z_1)T(z_2)...T(z_n)}\rangle_0$ is the coefficient of order $\epsilon_2\epsilon_3...\epsilon_n$ on the right-hand side of (\ref{deformed 1-pt expansion}). Applying this expansion to the equation of motion (\ref{EOM T}) and extracting the order $\epsilon_2\epsilon_3...\epsilon_n$, we obtain
\begin{align}
&-\partial_{\bar z_1}\langle{T(z_1)...T(z_n)}\rangle_0\notag\\
=&\sum_{i=2}^n\Big[\delta(z_1-z_i)\partial_{z_1}\langle{T(z_1)...T(z_{i-1})T(z_{i+1})...T(z_n)}\rangle_0\notag\\
&+2\partial_{z_1}\delta(z_1-z_i)\langle{T(z_1)...T(z_{i-1})T(z_{i+1})...T(z_n)}\rangle_0\Big]+\frac{k}{2}\delta_{n,2}\partial^3_{z_1}\delta(z_1-z_2), \label{EOM n-point}
\end{align}
which can be solved by applying the Green function on the torus,
\begin{align}
	&\langle{T(z_1)T(z_2)...T(z_n)}\rangle_0\notag\\
	=&\frac{1}{\text{Im}\tau}\int_{\text{T}^2}d^2v\langle{T(v)T(z_2)...T(z_n)}\rangle_0+\sum_{i=2}^n\Big[G(z_1-z_i)\partial_{z_i}\langle{T(z_2)...T(z_n)}\rangle_0\notag\\
	&-2\partial_{z_1}G(z_1-z_i)\langle{T(z_2)...T(z_n)}\rangle_0\Big]-\frac{k}{2}\delta_{n,2}\partial_{z_1}^3G(z_1-z_2). \label{recursion relation CS 1}
\end{align}
Plugging (\ref{65}) into (\ref{recursion relation CS 1}) we obtain
\begin{align}
	&\langle{T(z_1)T(z_2)...T(z_n)}\rangle_0\notag\\
	=&2\pi i\partial_{\tau}\langle{T(z_2)...T(z_n)}\rangle_0+\sum_{i=2}^n\Big[\Big(\zeta(z_1-z_i)-2\eta(z_1-z_i)\Big)\partial_{z_i}\langle{T(z_2)...T(z_n)}\rangle_0\notag\\
	&+2\Big(P(z_1-z_i)+2\eta\Big)\langle{T(z_2)...T(z_n)}\rangle_0\Big]+\frac{k}{2}\delta_{n,2}P''(z_1-z_2). \label{recursion relation CS 2}
\end{align}
We input the Brown-York tensor $T_{zz}$ in 3D gravity, which corresponds to one-point correlator $\langle{T}\rangle_0$ in boundary CFT. Then the multi-point correlators can be obtained by applying (\ref{recursion relation CS 2}) multiple times. For instance, the two-point correlator $\langle{TT}\rangle$ and the three-point correlator $\langle{TTT}\rangle$ of the Euclidean BTZ black hole take the forms
\begin{align}
	&\langle{T(z_1)T(z_2)}\rangle_0\notag\\
	=&2\pi i\partial_{\tau}\langle{T(z_2)}\rangle_0+2\Big(P(z_1-z_2)+2\eta\Big)\langle{T(z_2)}\rangle_0+\frac{k}{2}P''(z_1-z_2)\notag\\
	=&-\frac{2\pi^3ic}{3\tau^3}+\frac{\pi^2c}{3\tau^2}\Big(P(z_1-z_2)+2\eta\Big)+\frac{c}{12}P''(z_1-z_2), \\
	&\langle{T(z_1)T(z_2)T(z_3)}\rangle_0 \notag\\
	=&2\pi i\partial_{\tau}\langle{T(z_2)T(z_3)}\rangle_0+\sum_{i=2}^3\Big[\Big(\zeta(z_1-z_i)-2\eta(z_1-z_i)\Big)\partial_{z_i}\langle{T(z_2)T(z_3)}\rangle_0\notag\\
	&+2\Big(P(z_1-z_i)+2\eta\Big)\langle{T(z_2)T(z_3)}\rangle_0\Big]\notag\\
	=&-\frac{4\pi^4c}{\tau^4}-\frac{4\pi^3 ic}{3\tau^3}\Big(P(z_1-z_2)+P(z_1-z_3)+P(z_2-z_3)+6\eta\Big)\notag\\
	&+\frac{c}{12}\Big(2\pi i\partial_{\tau}+2P(z_1-z_2)+2P(z_1-z_3)+8\eta\Big)\Big(\frac{4\pi^2}{\tau^2}(P(z_2-z_3)+2\eta)+P''(z_2-z_3)\Big)\notag\\
	&+\Big(\zeta(z_1-z_2)-\zeta(z_1-z_3)+2\eta(z_2-z_3)\Big)\Big(\frac{4\pi^2}{\tau^2}P'(z_2-z_3)+P'''(z_2-z_3)\Big).
\end{align}

\section{Holographic correlators of stress tensor at a finite cutoff}\label{3}
There has been an intensive study on $T\bar{T}$ deformation \cite{Zamolodchikov:2004ce, Smirnov:2016lqw} of 2D quantum field theory in the past few years. The deformation is defined as a flow of the action in the direction of the $T\bar{T}$ operator
\begin{align} \label{TTbar deformation definition}
    \frac{d S^{[\mu]}}{d\mu} = \frac{1}{8}\int d^2x T\bar{T}^{[\mu]}
\end{align}
where $\mu$ is the continuous deformation parameter and the $T\bar{T}$ operator is a quadratic combination of the stress tensor
\begin{align} \label{TTbar operator definition}
    T\bar{T} = T^{ij} T_{ij} - {T^i_i}^2
\end{align}
As a solvable deformation, the $T\bar{T}$ deformation has many interesting properties, so it's natural to ask what is the holographic dual of the $T\bar{T}$ deformation of a holographic 2D CFT. It was proposed in \cite{McGough:2016lol} that the effect of $T\bar{T}$ deformation holographically corresponds to moving the conformal boundary to a finite cutoff surface $y=y_c$ in the AdS space, where $y$ is the Fefferman-Graham radial coordinate and the cutoff location $y_c$ is related to the deformation parameter $\mu$ by
\begin{align} \label{cutoff location deformation parameter relation}
    \mu = 16\pi G y_c^2
\end{align}
This proposal of cutoff AdS holography was generalized to higher dimensions in \cite{Taylor:2018xcy,Hartman:2018tkw,Shyam:2018sro,Belin:2020oib,Ebert:2024fpc}, and $\rm AdS_3$/$T\bar{T} \;\rm CFT_2$ was refined in \cite{Kraus:2022mnu} later. A natural prescription of the cutoff AdS holography is given by the generalized GKPW relation
\begin{align} \label{GKPW relation cutoff AdS holography}
    Z_{FT}[\gamma_{ij},J] = Z_{G}[\gamma_{ij},J] 
\end{align}
where the left-hand side is the field theory partition function with background metric $\gamma_{ij}$ and operator sources $J$, and the right-hand side is the gravity partition function with prescribed boundary conditions on the cutoff surface, the induced metric ${h}_{ij}$ and boundary value of fields $\phi_c$ satisfying
\begin{align} \label{source B.C. relation cutoff AdS holography}
    \gamma_{ij} &= y_c^2 {h}_{ij} \nonumber\\
    J &= y_c^{\Delta-d} \phi_c
\end{align}
An important perspective on the cutoff AdS holography is through the trace relation of $T\bar{T}$ deformed conformal field theory. In the large central charge limit it was derived in \cite{Guica:2019nzm} as the relation between one-point correlators of the stress tensor \footnote{The trace relation was also proved as the identity of classical fields for $T\bar{T}$-deformed CFT on curved space in \cite{Tolley:2019nmm}.}
\begin{align}
    \langle T^i_i \rangle = \frac{c}{24\pi} R[\gamma] + \frac{1}{4} \mu (\langle T^{ij}\rangle  \langle T_{ij}\rangle - {\langle T^i_i\rangle}^2)\label{one-point correlator TTBar trace relation}
\end{align}
With the holographic dictionary, the Brown-Henneaux central charge relation and (\ref{cutoff location deformation parameter relation}), we find this relation corresponds to the trace relation of the Brown-York tensor (\ref{BY tensor conservation}) at the cutoff $y=y_c$.

In this section, we will start with the trace relation to compute the holographic stress tensor correlators on a torus at a finite cutoff in the AdS space, extending the computation in the complex plane in \cite{Kraus:2018xrn,Li:2020pwa}. The original reasoning leading to the cutoff AdS/$T\bar{T}$ CFT correspondence only considers pure gravity, and the matter fields were only recently incorporated in \cite{Araujo-Regado:2022gvw}. We do not discuss the problem of a full cutoff AdS holography including matter fields in our paper.\footnote{For generic holographic correlation function of the scalar operators, one can refer to \cite{He:2023knl}. } Restricting ourselves to correlators of only the stress tensor, the computation based on the cutoff AdS picture is always valid as the holographic dual of $T\bar{T}$-deformed CFT.

\subsection{Holographic two-point correlators in cutoff BTZ black hole}
In this subsection, we compute the holographic stress tensor correlators on a cutoff torus in the Euclidean BTZ black hole spacetime, which presumably dominates in the high-temperature limit. The metric of rotating BTZ black hole can be written as \cite{Kraus:2006wn}
\begin{align}
    ds^2=&\frac{(r^2-r_+^2)(r^2-r_-^2)}{r^2}dt^2+\frac{r^2}{(r^2-r_+^2)(r^2-r_-^2)}dr^2+r^2(d\phi+i\frac{r_+r_-}{r^2}dt)^2,
\end{align}
where $r_+=R_+$ and $r_-=-iR_-$ for the outer horizon $R_+$ and the inner horizon $R_-$ of the black hole in the Lorentzian signature. The modular parameter takes the form
\begin{align}
    \tau=&\frac{i}{r_++r_-}
\end{align}
Now, let us return to the F-G coordinate system. Applying the following coordinate transformation
\begin{align}
    r=\sqrt{r_+^2+\frac{1}{4}(\frac{1}{\pi y}+\pi y(r_+^2-r_-^2))^2},\ \ \ \phi=\pi(z+\bar z),\ \ \ t=-i\pi(z-\bar z),
\end{align}
the bulk metric can be rewritten as 
\begin{align}
ds^2=&\frac{dy^2}{y^2}+\frac{1}{y^2}\Big[dzd\bar z-\pi^2y^2(\frac{1}{\tau^2}dz^2+\frac{1}{\bar\tau^2}d\bar z^2)+\pi^4y^4\frac{1}{\tau^2\bar\tau^2}dzd\bar z\Big].
\end{align}
To calculate the $T\bar T$-deformed correlators, we consider the boundary to be a hard radial cutoff at $y=y_c$. The deformed bulk solution is written as the Banados form \cite{Banados:1998gg},
\begin{align}
    ds^2=&\frac{dy^2}{y^2}+\frac{1}{y^2}\Big[dwd\bar w+y^2(\mathcal{L}_{y_c}dw^2+\bar{\mathcal{L}}_{y_c}d\bar w^2)+y^4\mathcal{L}_{y_c}\bar{\mathcal{L}}_{y_c}dwd\bar w\Big].
\end{align}
We use the following coordinate transformation,
\begin{align}
   z=w+y_c^2\bar{\mathcal{L}}_{y_c}\bar w,\ \ \ \bar z=y_c^2\mathcal{L}_{y_c}w+\bar w.
\end{align}
The periods of $z$ are $1$ and $\tau$. The new F-G coordinates $(y,z,\bar z)$ ensure that the boundary metric at $y_c$ takes on an Euclidean form. The $T\bar T$-deformed one-point correlators can be obtained from the bulk metric,
\begin{align}
\langle{T_{zz}}\rangle=&\frac{\mathcal{L}_{y_c}}{8\pi G(1-y_c^4\mathcal{L}_{y_c}\bar{\mathcal{L}}_{y_c})},\notag\\
\langle{T_{z\bar z}}\rangle=&\frac{2y_c^2\mathcal{L}_{y_c}\bar{\mathcal{L}}_{y_c}}{8\pi G(1-y_c^4\mathcal{L}_{y_c}\bar{\mathcal{L}}_{y_c})},\notag\\
\langle{T_{\bar z\bar z}}\rangle=&\frac{\bar{\mathcal{L}}_{y_c}}{8\pi G(1-y_c^4\mathcal{L}_{y_c}\bar{\mathcal{L}}_{y_c})}.
\end{align}
The next step is to find the relation between $(\mathcal{L}_{y_c},\bar{\mathcal{L}}_{y_c})$ and $(\mathcal{L}_{0},\bar{\mathcal{L}}_{0})=(-\frac{\pi^2}{\tau^2},-\frac{\pi^2}{\bar\tau^2})$. Following the prescription in \cite{Guica:2019nzm}, we proceed by equating the horizon area and angular momentum of the deformed black hole to those of the undeformed black hole, yielding
\begin{align}
    \mathcal{L}_{y_c}=\frac{\tau^2[(1+\frac{\pi^2y_c^2(\tau^2-\bar\tau^2)}{\tau^2\bar\tau^2})\sqrt{1+\frac{\pi^4y_c^4(\tau^2-\bar\tau^2)^2}{\tau^4\bar\tau^4}+\frac{2\pi^2y_c^2(\tau^2+\bar\tau^2)}{\tau^2\bar\tau^2}}-1-\frac{2\pi^2y_c^2}{\bar\tau^2}-\frac{\pi^4y_c^4(\tau^2-\bar\tau^2)^2}{\tau^4\bar\tau^4}]}{2\pi^2y_c^4}.
\end{align}
According to the generalized GKPW relation (\ref{GKPW relation cutoff AdS holography}), the higher-point correlators are computed by perturbing the boundary metric on the cutoff surface. Once again we substitute (\ref{perturbed boundary metric}) and (\ref{perturbed one-point function}) into conservation equation (\ref{one-point correlator conservation}) and $T\bar T$ trace relation (\ref{one-point correlator TTBar trace relation}). The terms of order $\epsilon^k$ read
\begin{align}
    \langle{T_{z\bar z}}\rangle^{[k]}=&\bar A\langle{T_{zz}}\rangle^{[k]}+A\langle{T_{\bar z\bar z}}\rangle^{[k]}+\mathcal{F}^{[k]}_{T_{z\bar z}},\label{general Tzzbar k}\\
    \partial_{\bar z}\langle{T_{zz}}\rangle^{[k]}=&-\partial_{z}\langle{T_{z\bar z}}\rangle^{[k]}+\mathcal{F}^{[k]}_{T_{zz}},\label{general Tzz k}\\
    \partial_{z}\langle{T_{\bar z\bar z}}\rangle^{[k]}=&-\partial_{\bar z}\langle{T_{z\bar z}}\rangle^{[k]}+\mathcal{F}^{[k]}_{T_{\bar z\bar z}}.\label{general Tzbarzbar k}
\end{align}
Here $\mathcal{F}^{[k]}_{T_{z\bar z}}$, $\mathcal{F}^{[k]}_{T_{zz}}$ and $\mathcal{F}^{[k]}_{T_{\bar z\bar z}}$ consist of lower-order coefficients and local functions of $f_{ij}$.  $A$ and $\bar A$ are constants determined by the one-point correlators,
\begin{align}
    A=&\frac{8\pi Gy^2_c\langle{T_{zz}}\rangle^{[0]}}{1+16\pi Gy^2_c\langle{T_{z\bar z}}\rangle^{[0]}}=\frac{y_c^2\mathcal{L}_{y_c}}{1+3y_c^4\mathcal{L}_{y_c}\bar{\mathcal{L}}_{y_c}},\notag\\
    \bar A=&\frac{8\pi Gy^2_c\langle{T_{\bar z\bar z}}\rangle^{[0]}}{1+16\pi Gy^2_c\langle{T_{z\bar z}}\rangle^{[0]}}=\frac{y_c^2\bar{\mathcal{L}}_{y_c}}{1+3y_c^4\mathcal{L}_{y_c}\bar{\mathcal{L}}_{y_c}}.
\end{align}
Plugging (\ref{general Tzzbar k}) into (\ref{general Tzz k}) and (\ref{general Tzbarzbar k}), and performing some straightforward calculations, we obtain
\begin{align}
    (\partial_z\partial_{\bar z}+\bar A\partial^2_{z}+A\partial^2_{\bar z})\langle{T_{zz}}\rangle^{[k]}=-\partial_z^2\mathcal{F}^{[k]}_{T_{z\bar z}}+(\partial_z+A\partial_{\bar z})\mathcal{F}^{[k]}_{T_{zz}}-A\partial_z\mathcal{F}^{[k]}_{T_{\bar z\bar z}}. \label{decoupled equation 1}
\end{align}
For simplicity, we consider the non-rotating BTZ black hole, where $r_-=0$ and $\tau$ is a purely imaginary number. We introduce the following coordinate transformation
\begin{align}\label{zZ transformation}
    z=&\frac{1}{2}\Big(1+\sqrt{\frac{1-2A}{1+2A}}\Big)Z+\frac{1}{2}\Big(1-\sqrt{\frac{1-2A}{1+2A}}\Big)\bar Z,\notag\\
    \bar z=&\frac{1}{2}\Big(1-\sqrt{\frac{1-2A}{1+2A}}\Big)Z+\frac{1}{2}\Big(1+\sqrt{\frac{1-2A}{1+2A}}\Big)\bar Z.
\end{align}
The periods in $Z$ are $1$ and $\Omega$ with the new modular parameter
\begin{align}
    \Omega=\sqrt{\frac{1+2A}{1-2A}}\tau.
\end{align}
The left-hand side of (\ref{decoupled equation 1}) can be written as
\begin{align}
     (\partial_z\partial_{\bar z}+\bar A\partial^2_{z}+A\partial^2_{\bar z})\langle{T_{zz}}\rangle^{[k]}=&(1+2A)\partial_Z\partial_{\bar Z}\langle{T_{zz}}\rangle^{[k]}.
\end{align}
Then equation (\ref{decoupled equation 1}) can be solved by the torus Green's function $\tilde G_{\Omega}(W-Z)$,
\begin{align}
    \frac{\delta^{k}\langle{T_{zz}(z)}\rangle^{[k]}}{\prod_{i=1}^k\delta f_{\alpha_i\beta_i}(z_i)}=&\int d^2W\Big(\frac{1}{\text{Im}\Omega}+\frac{1}{\pi}\tilde{G}_{\Omega}(W-Z)\partial_W\partial_{\bar W}\Big)\frac{\delta^{k+l}\langle{T_{zz}(w)}\rangle^{[k]}}{\prod_{i=1}^k\delta f_{\alpha_i\beta_i}(z_i)}. 
\end{align}
To the first-order in $\epsilon$, we have 
\begin{align}
    \langle{T_{z\bar z}}\rangle^{[1]}=&\bar A\langle{T_{zz}}\rangle^{[1]}+A\langle{T_{\bar z\bar z}}\rangle^{[1]}+\frac{1}{1+16\pi Gy_c^2\langle{T_{z\bar z}}\rangle^{[0]}}\Big[\Big(\langle{T_{\bar z\bar z}}\rangle^{[0]}-\frac{1}{16\pi G}\partial_{\bar z}^2\Big)f_{zz}\notag\\
    &-2\Big(\langle{T_{z\bar z}}\rangle^{[0]}-\frac{1}{16\pi G}\partial_z\partial_{\bar z}\Big)f_{z\bar z}+\Big(\langle{T_{zz}}\rangle^{[0]}-\frac{1}{16\pi G}\partial_z^2\Big)f_{\bar z\bar z}\Big],\notag\\
   \partial_Z\partial_{\bar Z}\langle{T_{zz}}\rangle^{[1]}=&\frac{1}{(1+2A)(1+16\pi Gy_c^2\langle{T_{z\bar z}}\rangle^{[0]})}\Big[\Big(8\pi Gy_c^2\langle{T_{\bar z\bar z}}\rangle^{[0]}[2\langle{T_{z\bar z}}\rangle^{[0]}-3\langle{T_{zz}}\rangle^{[0]}]\partial_z^2\notag\\
   &+2[\langle{T_{z\bar z}}\rangle^{[0]}+4\pi Gy_c^2(4(\langle{T_{z\bar z}}\rangle^{[0]})^2+\langle{T_{zz}}\rangle^{[0]}\langle{T_{\bar z\bar z}}\rangle^{[0]})]\partial_z\partial_{\bar z}+\frac{1}{16\pi G}\partial_z^2\partial_{\bar z}^2\notag\\
   &+16\pi Gy_c^2\langle{T_{zz}}\rangle^{[0]}\langle{T_{z\bar z}}\rangle^{[0]}\partial_{\bar z}^2\Big)f_{zz}+\Big(16\pi Gy_c^2\langle{T_{zz}}\rangle^{[0]}[2\langle{T_{\bar z\bar z}}\rangle^{[0]}-\langle{T_{z\bar z}}\rangle^{[0]}]\partial_z^2\notag\\
   &+16\pi Gy_c^2\langle{T_{zz}}\rangle^{[0]}\langle{T_{z\bar z}}\rangle^{[0]}\partial_z\partial_{\bar z}-\frac{1}{8\pi G}\partial_z^3\partial_{\bar z}\Big)f_{z\bar z}+\Big(\frac{1}{16\pi G}\partial_z^4+\notag\\
   &+2[\langle{T_{z\bar z}}\rangle^{[0]}-4\pi Gy_c^2\langle{T_{z\bar z}}\rangle^{[0]}(\langle{T_{zz}}\rangle^{[0]}-6\langle{T_{z\bar z}}\rangle^{[0]})]\partial_z^2\notag\\
   &-8\pi Gy_c^2\langle{T_{zz}}\rangle^{[0]}[2\langle{T_{z\bar z}}\rangle^{[0]}-3\langle{T_{zz}}\rangle^{[0]}]\partial_z\partial_{\bar z}\Big)f_{\bar z\bar z}\Big].\label{BTZ TTBar 2pt 0}
\end{align}
Using the torus Green's function $\tilde G_{\Omega}$, we obtain
\begin{align}
\frac{\delta\langle{T_{zz}(z)}\rangle}{\delta \gamma_{\alpha\beta}(z_1)}=&\frac{1}{\text{Im}\Omega}\int_{\text{T}^2} d^2W\frac{\delta\langle{T_{ww}(w)}\rangle}{\delta \gamma_{\alpha\beta}(z_1)}+\frac{1}{\pi(1+16\pi Gy_c^2\langle{T_{z\bar z}}\rangle)\sqrt{1-4A^2}}\notag\\
&\times\Big\lbrace\delta_{\alpha z}\delta_{\beta z}\Big[\frac{1}{16\pi G}\partial_{z_1}^2\partial_{\bar z_1}^2+8\pi Gy_c^2\langle{T_{\bar z\bar z}}\rangle[2\langle{T_{z\bar z}}\rangle-3\langle{T_{zz}}\rangle]\partial_{z_1}^2+2[\langle{T_{z\bar z}}\rangle\notag\\
&+4\pi Gy_c^2(4\langle{T_{z\bar z}}\rangle^2+\langle{T_{zz}}\rangle\langle{T_{\bar z\bar z}}\rangle)]\partial_{z_1}\partial_{\bar z_1}+16\pi Gy_c^2\langle{T_{zz}}\rangle\langle{T_{z\bar z}}\rangle\partial_{\bar z_1}^2\Big]\notag\\
&+\delta_{\alpha z}\delta_{\beta \bar z}\Big[8\pi Gy_c^2\langle{T_{zz}}\rangle[2\langle{T_{\bar z\bar z}}\rangle-\langle{T_{z\bar z}}\rangle]\partial_{z_1}^2+8\pi Gy_c^2\langle{T_{zz}}\rangle\langle{T_{z\bar z}}\rangle\partial_{z_1}\partial_{\bar z_1}\notag\\
&-\frac{1}{16\pi G}\partial_{z_1}^3\partial_{\bar z_1}\Big]+\delta_{\alpha \bar z}\delta_{\beta \bar z}\Big[\frac{1}{16\pi G}\partial_{z_1}^4+2\langle{T_{z\bar z}}\rangle[1-4\pi Gy_c^2(\langle{T_{zz}}\rangle-6\langle{T_{z\bar z}}\rangle)]\partial_{z_1}^2\notag\\
   &-8\pi Gy_c^2\langle{T_{zz}}\rangle[2\langle{T_{z\bar z}}\rangle-3\langle{T_{zz}}\rangle]\partial_{z_1}\partial_{\bar z_1}\Big]\Big\rbrace\tilde{G}_{\Omega}(Z_1-Z).\label{BTZ TTBar 2pt 1}
\end{align}
The first term on the right-hand side is a one-point-averaged correlator, which can be computed utilizing the technique described in subsection \ref{2.2}:
\begin{align}
        (\bar{\tau}-\tau) \partial_{\tau} \langle{T_{\alpha\beta}}\rangle &=\mathcal{L}_{(z_1-\bar{z}_1)\partial_{z_1}}\langle{T_{\alpha\beta}}\rangle
    +\int_{\text{T}^2} d^2w \Big( \frac{\delta\langle{T_{\alpha\beta}(z_1)}\rangle}{\delta \gamma_{\bar{w}\bar{w}}(w)} -  \frac{\delta\langle{T_{\alpha\beta}(z_1)}\rangle}{\delta \gamma_{w\bar{w}}(w)} \Big),\notag\\
    (\tau-\bar{\tau}) \partial_{\bar{\tau}} \langle{T_{\alpha\beta}}\rangle &= \mathcal{L}_{(\bar{z}_1-z_1)\partial_{\bar{z}_1}}\langle{T_{\alpha\beta}}\rangle
    + \int_{\text{T}^2} d^2w \Big( \frac{\delta\langle{T_{\alpha\beta}(z_1)}\rangle}{\delta \gamma_{ww}(w)} -  \frac{\delta\langle{T_{\alpha\beta}(z_1)}\rangle}{\delta \gamma_{w\bar{w}}(w)}\Big). \label{}
\end{align}
Note that the term corresponds to the global Weyl transformation does not vanish any longer. However, it can be figured out with the help of the fact that the variation with respect to $\gamma_{z\bar{z}}$ can be transformed into the one with respect to $\gamma_{zz}$ and $\gamma_{\bar{z}\bar{z}}$, e.g.
\begin{align}
    \frac{\delta \langle{T_{zz}(z_1)}\rangle}{\delta{\gamma_{w\bar w}(w)}} = \frac{\delta \langle{T_{w\bar w}(w)}\rangle}{\delta \gamma_{\bar{z}\bar{z}}(z_1)} -\delta(z_1-w) \langle{T_{zz}}\rangle,
\end{align}
and that $\frac{\delta \langle{T_{w\bar w}(w)}\rangle}{\delta \gamma_{\bar{z}\bar{z}}(z_1)}$ can be expressed in terms of $\frac{\delta \langle{T_{ww}(w)}\rangle}{\delta \gamma_{\bar{z}\bar{z}}(z_1)}$, $\frac{\delta \langle{T_{\bar w\bar w}(w)}\rangle}{\delta \gamma_{\bar{z}\bar{z}}(z_1)}$ and contact terms. After a simple calculation, we find
\begin{align}
    \int_{\text{T}^2} d^2w\frac{\delta\langle{T_{ww}(w)}\rangle}{\delta \gamma_{zz}(z_1)}=&\frac{-2}{1-2A}\Big[i\text{Im}\tau[A\partial_{\tau}-(1-A)\partial_{\bar\tau}]+A\Big]\langle{T_{zz}}\rangle+\frac{2\langle{T_{z\bar z}}\rangle}{1-2A},\notag\\
    \int_{\text{T}^2} d^2w\frac{\delta\langle{T_{ww}(w)}\rangle}{\delta \gamma_{\bar z\bar z}(z_1)}=&\frac{-2}{1-2A}\Big[i\text{Im}\tau[(1-A)\partial_{\tau}-A\partial_{\bar\tau}]+1+A\Big]\langle{T_{zz}}\rangle,\notag\\
    \int_{\text{T}^2} d^2w\frac{\delta\langle{T_{ww}(w)}\rangle}{\delta \gamma_{z\bar z}(z_1)}=&A\int_{\text{T}^2} d^2w\Big[\frac{\delta\langle{T_{ww}(w)}\rangle}{\delta \gamma_{\bar z\bar z}(z_1)}+\frac{\delta\langle{T_{ww}(w)}\rangle}{\delta \gamma_{zz}(z_1)}\Big]-2A\langle{T_{z\bar z}}\rangle.\label{BTZ TTBar Constant}
\end{align}
By combining (\ref{BTZ TTBar 2pt 0})(\ref{BTZ TTBar 2pt 1})(\ref{BTZ TTBar Constant}), we obtain
\begin{align}
    \langle{T_{zz}(z)T_{zz}(z_1)}\rangle=&\frac{[-i(1-A)\partial_{\tau}+iA\partial_{\bar\tau}-\frac{1+A}{\text{Im}\tau}]\langle{T_{zz}}\rangle}{1-2A}+\frac{1}{2\pi(1+16\pi Gy_c^2\langle{T_{z\bar z}}\rangle)\sqrt{1-4A^2}}\notag\\
    &\times\Big[\frac{1}{16\pi G}\partial_{z_1}^4+2\langle{T_{z\bar z}}\rangle[1-4\pi Gy_c^2(\langle{T_{zz}}\rangle-6\langle{T_{z\bar z}}\rangle)]\partial_{z_1}^2\notag\\
   &-8\pi Gy_c^2\langle{T_{zz}}\rangle[2\langle{T_{z\bar z}}\rangle-3\langle{T_{zz}}\rangle]\partial_{z_1}\partial_{\bar z_1}\Big]\tilde{G}_{\Omega}(Z_1-Z),\notag\\
    \langle{T_{zz}(z)T_{\bar z\bar z}(z_1)}\rangle=&\frac{[-iA\partial_{\tau}-i(1-A)\partial_{\bar\tau}-\frac{A}{\text{Im}\tau}]\langle{T_{zz}}\rangle+\frac{1}{\text{Im}\tau}\langle{T_{z\bar z}}\rangle}{1-2A}+\frac{1}{2\pi(1+16\pi Gy_c^2\langle{T_{z\bar z}}\rangle)\sqrt{1-4A^2}}\notag\\
    &\times \Big[\frac{1}{16\pi G}\partial_{z_1}^2\partial_{\bar z_1}^2+8\pi Gy_c^2\langle{T_{\bar z\bar z}}\rangle[2\langle{T_{z\bar z}}\rangle-3\langle{T_{zz}}\rangle]\partial_{z_1}^2+16\pi Gy_c^2\langle{T_{zz}}\rangle\langle{T_{z\bar z}}\rangle\partial_{\bar z_1}^2\notag\\
&+2[\langle{T_{z\bar z}}\rangle+4\pi Gy_c^2(4\langle{T_{z\bar z}}\rangle^2+\langle{T_{zz}}\rangle\langle{T_{\bar z\bar z}}\rangle)]\partial_{z_1}\partial_{\bar z_1}\Big]\tilde{G}_{\Omega}(Z_1-Z),\notag\\
    \langle{T_{z\bar z}(z)T_{zz}(z_1)}\rangle=&A(\langle{T_{zz}(z)T_{zz}(z_1)}\rangle+\langle{T_{\bar z\bar z}(z)T_{zz}(z_1)}\rangle)\notag\\
    &+\frac{1}{2(1+16\pi Gy^2_c\langle{T_{z\bar z}}\rangle)}(\langle{T_{zz}}\rangle-\frac{1}{16\pi G}\partial_{z}^2)\delta(z-z_1),\notag\\
     \langle{T_{z\bar z}(z)T_{z\bar z}(z_1)}\rangle=&A(\langle{T_{zz}(z)T_{z\bar z}(z_1)}\rangle+\langle{T_{\bar z\bar z}(z)T_{z\bar z}(z_1)}\rangle)\notag\\
     &-\frac{1}{2(1+16\pi Gy^2_c\langle{T_{z\bar z}}\rangle)}(\langle{T_{z\bar z}}\rangle-\frac{1}{16\pi G}\partial_z\partial_{\bar z})\delta(z-z_1),\notag\\
         \langle{T_{\bar z\bar z}(z)T_{\bar z\bar z}(z_1)}\rangle=&\text{ c.c. of }\langle{T_{zz}(z)T_{zz}(z_1)}\rangle,\notag\\
          \langle{T_{z\bar z}(z)T_{\bar z\bar z}(z_1)}\rangle=&\text{ c.c. of }\langle{T_{z\bar z}(z)T_{zz}(z_1)}. \label{general two-point stress tensor}
\end{align}
\subsection{Recursion relation of correlators}

Consider the variation of the form $\delta \gamma_{zz}=\epsilon f_{zz}$, $\delta\gamma_{\bar{z}\bar{z}} =\bar{\epsilon} f_{\bar{z}\bar{z}}$,  $\delta\gamma_{z\bar{z}}=\epsilon' f_{z\bar{z}}$, and then the terms of order $\epsilon^l \bar{\epsilon}^m \epsilon'^n$ in the conservation equation and the trace relation read
\begin{align}
    {T_{z\bar z}}^{[l,m,n]}=&\bar A {T_{zz}}^{[l,m,n]}+A{T_{\bar z\bar z}}^{[l,m,n]}+\mathcal{F}^{[l,m,n]}_{T_{z\bar z}},\label{general Tzzbar l,m,n}\\
    \partial_{\bar z}{T_{zz}}^{[l,m,n]}=&-\partial_{z}{T_{z\bar z}}^{[l,m,n]}+\mathcal{F}^{[l,m,n]}_{T_{zz}},\label{general Tzz l,m,n}\\
    \partial_{z}{T_{\bar z\bar z}}^{[l,m,n]}=&-\partial_{\bar z}{T_{z\bar z}}^{[l,m,n]}+\mathcal{F}^{[l,m,n]}_{T_{\bar z\bar z}},\label{general Tzbarzbar l,m,n}
\end{align}
where $\mathcal{F}^{[l,m,n]}_{T_{z\bar z}}$, $\mathcal{F}^{[l,m,n]}_{T_{zz}}$ and $\mathcal{F}^{[l,m,n]}_{T_{\bar z\bar z}}$
consist of lower-order coefficients and local functions of $f_{ij}$. $A$ and $\bar A$ are constants defined as before. By eliminating $T_{z\bar z}^{[l,m,n]}$, we obtain two second order differential equation on $T_{zz}^{[l,m,n]}$ and $T_{\bar{z}\bar{z}}^{[l,m,n]}$
\begin{align}
    \partial_Z\partial_{\bar Z} T_{zz}^{[l,m,n]}=&\frac{1}{1+2A}\big( -\partial^2 \mathcal{F}^{[l,m,n]}_{T_{z\bar z}} +(\partial +A \bar{\partial}) \mathcal{F}^{[l,m,n]}_{T_{zz}} -A \partial \mathcal{F}^{[l,m,n]}_{T_{\bar z\bar z}} \big), \nonumber\\
    \partial_Z\partial_{\bar Z} T_{\bar{z}\bar{z}}^{[l,m,n]}=&\frac{1}{1+2\bar{A}} \big( -\bar{\partial}^2 \mathcal{F}^{[l,m,n]}_{T_{z\bar z}} -\bar{A} \bar{\partial} \mathcal{F}^{[l,m,n]}_{T_{zz}} +(\bar{A} \partial +\bar{\partial}) \mathcal{F}^{[l,m,n]}_{T_{\bar z\bar z}} \big),
\end{align}
with $Z,\bar{Z}$ defined by (\ref{zZ transformation}). Then, as before, $T_{zz}^{[l,m,n]}$ and $T_{\bar{z}\bar{z}}^{[l,m,n]}$ can be solved by using the following identity,
\begin{align}\label{solve T lmn by the torus Green's function}
    T_{ij}^{[l,m,n]}(z)=&\int d^2W\Big(\frac{1}{\text{Im}\Omega}+\frac{1}{\pi}\tilde{G}_{\Omega}(W-Z)\partial_W\partial_{\bar W}\Big)T_{ij}^{[l,m,n]}(w). 
\end{align}
The first term on the right-hand side can be computed by setting $O=T_{zz}^{[l,m-1,n]}$ in (\ref{global metric variation and modular differentiation eqn 1}) and (\ref{global metric variation and modular differentiation eqn 2}),
\begin{align}
    (\bar{\tau}-\tau) \partial_{\tau} T_{zz}^{[l,m-1,n]} &=\mathcal{L}_{(z-\bar{z})\partial_{z}} T_{zz}^{[l,m-1,n]} 
    +\int_{\text{T}^2} d^2w \Big( \frac{\delta T_{zz}^{[l,m,n]}(z)}{\delta f_{\bar{w}\bar{w}}(w)} -  \frac{\delta T_{zz}^{[l,m-1,n+1]}(z)}{\delta f_{w\bar{w}}(w)} \Big),\notag\\
    (\tau-\bar{\tau}) \partial_{\bar{\tau}} T_{zz}^{[l,m-1,n]} &= \mathcal{L}_{(\bar{z}-z)\partial_{\bar{z}}} T_{zz}^{[l,m-1,n]}
    + \int_{\text{T}^2} d^2w \Big( \frac{\delta T_{zz}^{[l+1,m-1,n]}(z)}{\delta f_{ww}(w)} -  \frac{\delta T_{zz}^{[l,m-1,n+1]}(z)}{\delta f_{w\bar{w}}(w)}\Big). \label{global metric variation and modular differentiation eqn for TTbar recursion relation}
\end{align}
The terms correspond to the global Weyl transformation can be transformed into the variation with respect to $f_{zz}$ and $f_{\bar{z}\bar{z}}$ by the following relation,
\begin{gather}
    \frac{\delta T_{zz}^{[l,m,n]}(z)}{\delta{f_{\bar{w} \bar w}(w)}} = \frac{\delta T_{ww}^{[l,m,n]}(w)}{\delta f_{\bar{z}\bar{z}}(z)}, \quad
    \frac{\delta T_{zz}^{[l+1,m-1,n]}(z)}{\delta{f_{ww}(w)}} = \frac{\delta T_{\bar{w}\bar{w}}^{[l,m,n]}(w)}{\delta f_{\bar{z}\bar{z}}(z)}, \nonumber\\
    \frac{\delta T_{zz}^{[l,m-1,n+1]}(z)}{\delta{f_{w\bar w}(w)}} = \frac{\delta T_{w\bar{w}}^{[l,m,n]}(w)}{\delta f_{\bar{z}\bar{z}}(z)} -\delta(z-w) T_{zz}^{[l,m-1,n]}.\label{changing the position of variation}
\end{gather}
With the help of (\ref{changing the position of variation}) and (\ref{general Tzzbar l,m,n}), (\ref{global metric variation and modular differentiation eqn for TTbar recursion relation}) becomes 
\begin{align}
    &(\bar{\tau}-\tau) \partial_{\tau} T_{zz}^{[l,m-1,n]} =\mathcal{L}_{(z-\bar{z})\partial_{z}} T_{zz}^{[l,m-1,n]} \nonumber\\
    &+\int_{\text{T}^2} d^2w \Big( (1-\bar{A}) \frac{\delta T_{ww}^{[l,m,n]}(w)}{\delta f_{\bar{z}\bar{z}}(z)} -  A \frac{\delta T_{\bar{w}\bar{w}}^{[l,m,n]}(w)}{\delta f_{\bar{z}\bar{z}}(z)} -\frac{\delta \mathcal{F}_{T_{w\bar{w}}}^{[l,m,n]}(w)}{\delta f_{\bar{z}\bar{z}}(z)} +\delta(z-w) T_{zz}^{[l,m-1,n]}(z) \Big),\\
    &(\tau-\bar{\tau}) \partial_{\bar{\tau}} T_{zz}^{[l,m-1,n]} = \mathcal{L}_{(\bar{z}-z)\partial_{\bar{z}}} T_{zz}^{[l,m-1,n]} \nonumber\\
    & +\int_{\text{T}^2} d^2w \Big( -\bar{A} \frac{\delta T_{ww}^{[l,m,n]}(w)}{\delta f_{\bar{z}\bar{z}}(z)} + ( 1- A ) \frac{\delta T_{\bar{w}\bar{w}}^{[l,m,n]}(w)}{\delta f_{\bar{z}\bar{z}}(z)} -\frac{\delta \mathcal{F}_{T_{w\bar{w}}}^{[l,m,n]}(w)}{\delta f_{\bar{z}\bar{z}}(z)} +\delta(z-w) T_{zz}^{[l,m-1,n]}(z) \Big) . \label{}
\end{align}
Solve these two equations, and we obtain 
\begin{align}
    \int_{\text{T}^2} d^2w \frac{\delta T_{ww}^{[l,m,n]}(w)}{\delta f_{\bar{z}\bar{z}}(z)} = \frac{(1-A) a^{[l,m,n]} +A \bar{a}^{[l,m,n]}-b^{[l,m,n]}}{A- \bar{A}-1}, \\
    \int_{\text{T}^2} d^2w \frac{\delta T_{\bar{w}\bar{w}}^{[l,m,n]}(w)}{\delta f_{\bar{z}\bar{z}}(z)} = \frac{\bar{A} a^{[l,m,n]} + (1-\bar{A})\bar{a}^{[l,m,n]}-b^{[l,m,n]}}{A- \bar{A}-1} ,
\end{align}
where $a^{[l,m,n]}$, $\bar{a}^{[l,m,n]}$ and $b^{[l,m,n]}$ also consist of lower-order coefficients and local functions of $f_{ij}$, i.e.
\begin{align}
    a^{[l,m,n]} = (\bar{\tau}-\tau) \partial_{\tau} T_{zz}^{[l,m-1,n]} - \mathcal{L}_{(z-\bar{z})\partial_{z}} T_{zz}^{[l,m-1,n]}, \\
    \bar{a}^{[l,m,n]} = (\tau-\bar{\tau}) \partial_{\bar{\tau}} T_{zz}^{[l,m-1,n]} - \mathcal{L}_{(\bar{z}-z)\partial_{\bar{z}}} T_{zz}^{[l,m-1,n]}, \\
    b^{[l,m,n]} =T_{zz}^{[l,m-1,n]} - \int_{\text{T}^2} d^2w \frac{\delta \mathcal{F}_{T_{w\bar{w}}}^{[l,m,n]}(w)}{\delta \gamma_{\bar{z}\bar{z}}(z)}. 
\end{align}
Making use of the identity
\begin{align}
    \int d^2 w T_{ij}^{[l,m,n]}(w)=&\int d^2 z f_{\bar{z}\bar{z}}(z) \frac{\delta}{\delta f_{\bar{z}\bar{z}}(z)} \int d^2 w T_{ij}^{[l,m,n]}(w), 
\end{align}
we have
\begin{align}\label{one-point averaged T lmn}
    \int d^2 w T_{ww}^{[l,m,n]}(w)=&\int d^2 z f_{\bar{z}\bar{z}} \frac{(1-A) a^{[l,m,n]} +A \bar{a}^{[l,m,n]}-b^{[l,m,n]}}{A- \bar{A}-1}.
\end{align}
Combining (\ref{one-point averaged T lmn}) with (\ref{solve T lmn by the torus Green's function}), we obtain the full form of $T_{ij}^{[l,m,n]}$. For example, the $zz$-component is 
\begin{align}\label{TTbar recur}
    T_{zz}^{[l,m,n]}(z)=&\int d^2 z f_{\bar{z}\bar{z}} \frac{(1-A) a^{[l,m,n]} +A \bar{a}^{[l,m,n]}-b^{[l,m,n]}}{A- \bar{A}-1} \notag\\
    &+\frac{1}{\pi(1+2A)}\int d^2W\tilde{G}_{\Omega}(W-Z) \big( -\partial^2 \mathcal{F}^{[l,m,n]}_{T_{z\bar z}} +(\partial +A \bar{\partial}) \mathcal{F}^{[l,m,n]}_{T_{zz}} -A \partial \mathcal{F}^{[l,m,n]}_{T_{\bar z\bar z}} \big).
\end{align}
By taking $(l+m+n)$ times of functional derivatives with respect to $f_{zz}^l f_{\bar{z}\bar{z}}^m f_{z\bar{z}}^n$ in (\ref{TTbar recur}), one can obtain the recursion relation of the stress tensor correlators.

\section{Correlators of KdV charges in thermal AdS}
A direct application of the stress tensor correlators is that they can be used to represent the KdV charges in CFT. In this section, we utilize the holographic stress tensor correlators in thermal AdS$_3$ to compute the correlators of KdV charges in the dual CFT. Using the flow equation introduced by the authors of \cite{LeFloch:2019wlf}, we obtain the correlators of KdV charges in the $T\bar T$-deformed CFT.

\subsection{In low-temperature phase of CFT}
We start by reviewing some basics of KdV charges. Consider a cylinder with coordinate $\{t,x\}$ circumference $L$, i.e. 
$z=x +i t$, $x\sim x+L$. 
The theory is a CFT with a Virasoro algebra
\begin{align}
    T(\sigma)=-\frac{c}{24}+ \sum_{n=-\infty}^{+\infty}L_{-n}e^{in\sigma},
\end{align}
\begin{align}
    [L_m,L_n]=(n-m)L_{n+m}+\frac{c}{12}(n^3-n)\delta_{m+n,0}.
\end{align}
There exists an infinite set of commuting Integrals of Motion (IM) generated by various “composite fields” built as a power of $T(\sigma)$ and its derivatives which have the form \cite{Kupershmidt:1989bf,Bazhanov:1994ft}
\begin{align}
    P_{s}=\frac{1}{2\pi}\int_{0}^{L}dz T_{s+1}(z),
\end{align}
where the densities $T_{s+1}(u)$ are appropriately regularized polynomials in $T(u)$ and subscript $s$ takes an odd value. The first few densities $T_{s+1}(u)$ can be written as 
\begin{align}
    T_2(z)=T(z),\quad T_4(z)=:T^2(z):,\quad T_6(z)= :T^3(z): +\frac{c+2}{12}:(T'(z))^2:.
\end{align}
Here the prime stands for the derivative and $: :$ denotes appropriately regularized products of the fields, for example
\begin{align}
    :T^2(u):=\oint_{\mathcal{C}_u} \frac{dw}{2\pi i}\frac{1}{w-u} \mathcal{T}[T(w)T(u)],
\end{align}
where the symbol $\mathcal{T}$ denotes the “chronological ordering”, i.e.
\begin{align}
    \mathcal{T}[A(w)B(u)]=
    \left\{
    \begin{aligned}
        A(w)B(u),\quad \text{Im}u>\text{Im}w;\\
        B(u)A(w),\quad \text{Im}w>\text{Im}u.
    \end{aligned}
    \right.
\end{align}
Consider the correlator of the first KdV charge $\langle{P_1}\rangle$, which can be written in terms of the stress tensor correlator $\langle{T}\rangle$ on the cylinder. We employ the one-point correlator in the thermal AdS background and let the modular parameter go to $i\infty$,
\begin{align}
    \langle P_{1}\rangle = \frac{1}{2\pi}\int_{0}^{L}dz \langle T(z) \rangle_{\text{cyl}.}= -\frac{1}{L}\lim_{\tau\to i\infty}\int_{0}^{1}dz \langle T_{zz}(z) \rangle_{\text{AdS}}= \frac{\pi}{8GL}.\label{136}
\end{align}
The correlator of the second KdV charge is
\begin{align}
    \langle P_{3}\rangle = &\frac{1}{2\pi}\int_{0}^{L}dz \langle :T^2(z): \rangle_{\text{cyl}.} = \frac{2\pi}{L^3} \lim_{\tau\to i\infty}\int_{0}^{1}dz \oint_{\mathcal{C}_z} \frac{dw}{2\pi i}\frac{1}{w-z} \langle T_{zz}(w)T_{zz}(z) \rangle_{\text{AdS}}. \label{137}
\end{align}
The two-point correlator $\langle{T_{zz}T_{zz}}\rangle$ in thermal AdS$_3$ has been obtained in our previous work \cite{He:2023hoj},
\begin{equation}
    \langle T_{zz}(z) T_{zz}(w)\rangle_{\text{AdS}} = \frac{1}{16\pi G}[-\frac{1}{2\pi}\partial_z^3 G_\tau(z-w) - 2\pi \partial_z G_\tau(z-w) + \frac{2\pi^2}{\text{Im}\tau}].\label{138}
\end{equation}
Plugging (\ref{138}) into (\ref{137}), we obtain
\begin{align}
    &\langle P_{3}\rangle=\frac{\pi}{2GL^3} \lim_{\tau\to i\infty} \zeta_{\tau}(\frac{1}{2}).
\end{align}
By employing the two-point correlator $\langle{T_{zz}T_{zz}}\rangle$, we can also compute the two-point function of $P_1$,
\begin{align}
    \langle P_1 P_1 \rangle =& \frac{1}{L^2} \lim_{\tau\to i\infty} \int_{0}^{1}dz \int_{0}^{1}dw \langle T(z) T(w) \rangle_{\text{AdS}} = \frac{\pi}{8 G L^2} \lim_{\tau\to i\infty}  \frac{1}{\text{Im}\tau} =0. 
\end{align}

\subsection{In low temperature phase of $T\bar{T}$ deformed theory}
In the $T\bar T$-deformed theory, the conservation equation is 
\begin{align}
    \partial_{\bar z}T^{\lambda}_{s+1} = \partial_z \Theta^{\lambda}_{s-1},
\end{align} 
and the KdV charges can be adjusted to remain conserved after the deformation, i.e.
\begin{align}
    P_{s}^\lambda=\frac{1}{2\pi}\int_{0}^{L}dz T^\lambda_{s+1}(z) + d\bar{z} \Theta^\lambda_{s-1},
\end{align}
where $\lambda$ is the deformation parameter. The flow equation of $\langle P^\lambda_s \rangle_n$ is \cite{LeFloch:2019wlf}
\begin{align}\label{Ps flow}
    \partial_{\lambda} \langle P^\lambda_s \rangle_n =-\pi^2 (E_n^\lambda \partial_L \langle P^\lambda_s \rangle_n +P_n^\lambda \frac{s \langle P^\lambda_s \rangle_n }{L} ),
\end{align}
where $E_n^\lambda$ and $P_n^\lambda$ is the eigenvalue of the Hamiltonian and the momentum according to $|n\rangle^\lambda$, which is given by solving the Burgers equation \cite{Smirnov:2016lqw,Cavaglia:2016oda,Zamolodchikov:2004ce},
\begin{equation}
    E_n^\lambda=\frac{L}{2\lambda} \Big( \sqrt{1 + \frac{4\lambda E_n}{L} + \frac{4\lambda^2(P_n)^2}{L^2} } -1 \Big), \quad P_n^\lambda=P_n.
\end{equation}
By solving (\ref{Ps flow}) with the initial condition 
\begin{equation}\label{P_s^0}
    \langle P_s^{(0)}(L) \rangle_n = \frac{p_s}{L^{|s|}},
\end{equation}
where $p_s$ are some numbers only depending on the state, but not on $L$, we get the $T\bar T$-deformed $\langle P_s^\lambda(L) \rangle_n$ \cite{LeFloch:2019wlf},
\begin{align}
    \langle P_s^\lambda(L) \rangle_n = 
    \left\{ 
    \begin{aligned}
        &\frac{p_s}{(p_1)^s} (\langle P_1^\lambda(L) \rangle_n)^s, \qquad s>0; \\
        &\frac{p_s}{(p_{-1})^{|s|}} (\langle P_{-1}^\lambda(L) \rangle_n)^s, \quad s<0.
    \end{aligned}
    \right.
\end{align}
Here $P_{-1}$ denotes the complex conjugate of $P_{1}$. It means that we are able to calculate $\langle P_s^\lambda \rangle_n$ if we know $\langle P_s^{(0)} \rangle_n$ and $\langle P_{\pm 1}^\lambda \rangle_n$. $\langle P_{1}^\lambda \rangle_n$ can be represented by deformed energy and momentum 
\begin{align}
    \langle P_{1}^\lambda \rangle_n =&-\frac{E_n^\lambda+P_n^\lambda}{2} \notag\\
    =&\frac{L}{4\lambda} \Big(1 - \sqrt{1 + \frac{4\lambda E_n^{(0)}}{L} + \frac{4\lambda^2(P_n^{(0)})^2}{L^2} } \Big) -\frac{P_n^{(0)}}{2} \notag\\
    =&\frac{L}{4\lambda} \Big(1 - \sqrt{1 - \frac{4\lambda (\langle P_1^{(0)} \rangle_n +\langle P_{-1}^{(0)} \rangle_n)}{L} + \frac{4\lambda^2(\langle P_1^{(0)} \rangle_n -\langle P_{-1}^{(0)} \rangle_n)^2}{L^2} } \Big) +\frac{(\langle P_1^{(0)} \rangle_n -\langle P_{-1}^{(0)} \rangle_n)}{2}. 
\end{align}
where in the third line we have used 
\begin{align}
    E_n^{(0)}=&- (\langle P_1^{(0)} \rangle_n +\langle P_{-1}^{(0)} \rangle_n),\\
    P_n^{(0)}=&- (\langle P_1^{(0)} \rangle_n -\langle P_{-1}^{(0)} \rangle_n). \label{148}
\end{align}
Now let us calculate the deformed one-point correlators on the vacuum state. Plugging (\ref{136}) into (\ref{148}), we obtain\footnote{We note that the result for $\langle{P_{1}^{\lambda}}\rangle$ here align with that in \cite{He:2020cxp}, where the calculation was conducted within the framework of free bosonic theory. This observation may be coincidental; however, further comparisons with other KdV charges might yield contrasting results.} 
\begin{align}\label{P_1^lambda}
    \langle P_{1}^\lambda \rangle_0 =&\frac{L}{4\lambda} \Big(1 - \sqrt{1 - \frac{\pi\lambda }{GL^2} } \Big)= \frac{\pi}{8G} +\frac{\pi^2 \lambda}{32G^2 L^3} +\ldots .
\end{align}

In order to obtain $\langle P_{3}^\lambda \rangle$, note first that 
\begin{align}
    \langle P_{1}\rangle=\frac{\pi}{8GL},\quad \langle P_{3}\rangle =\frac{\pi}{2GL^3} \lim_{\tau\to i\infty} \zeta_{\tau}(\frac{1}{2}),
\end{align}
and we see that 
\begin{align}
    p_1=\frac{\pi}{8G},\quad p_3=\frac{\pi}{2G} \lim_{\tau\to i\infty} \zeta_{\tau}(\frac{1}{2}).
\end{align}
Then
\begin{align}
    \langle P_{3}^\lambda \rangle =& \frac{p_3}{(p_1)^3}(\langle P_{1}^\lambda \rangle)^3 \notag\\
    =&\frac{4 G^2 L^3}{\pi^2 \lambda^3} \lim_{\tau\to i\infty} \zeta_{\tau}(\frac{1}{2}) \Big(1 - \sqrt{1 - \frac{\pi\lambda }{GL^2} } \Big)^3 \notag \\
    =&\lim_{\tau\to i\infty} \zeta_{\tau}(\frac{1}{2})(\frac{\pi}{2GL^3} +\frac{3\pi^2 \lambda}{8G^2 L^5} +\ldots).
\end{align}

\section{Conclusions and perspectives}\label{4}
This paper investigates stress tensor correlators within the holographic framework, particularly focusing on their computation on manifolds with nontrivial topology. While conventional methods like GKPW prescriptions suffice for simple cases, addressing higher point correlators necessitates addressing diffeomorphism invariance and gauge fixing intricacies. Here, we present an algorithm tailored for computing generic stress tensor correlators on the torus in the AdS/CFT context. Understanding the ensemble average of CFTs/AdS duality aids in analyzing holographic correlators in conformal field theories on nontrivial topologies. We employ the dominant gravitational saddle in AdS$_3$ and utilize both metric and Chern-Simons formulations to compute correlators. Recurrence relations among correlators emerge, mirroring those established in pure conformal field theory on the torus \cite{He:2020udl}. Additionally, we explore lower-point correlators of KdV charges at low temperatures, showcasing the practical utility of our theoretical framework in elucidating modifications induced by $T\bar{T}$ deformations in a low-temperature regime.

In our methodology, stress tensor correlators are evaluated by perturbing the spin-2 bulk field around the AdS background. This algorithm can be extended to compute correlators of higher-point generic operators, corresponding holographically to diverse spin-bulk fields such as scalar operators as demonstrated by \cite{He:2023knl}. Moreover, an intriguing avenue for future exploration is extending the algorithm to correlators on higher genus Riemann surfaces, crucial for comprehending the ensemble average of dual CFTs. Progress in this direction will be reported in forthcoming works.


\section*{Acknowledgments}
We want to thank Bin Chen, Alex Maloney, Cheng Peng, and Xi-Nan Zhou for useful discussions related to this work. S.H. also would like to appreciate the financial support from the Max Planck Partner Group, the Fundamental Research Funds for the Central Universities, and the Natural Science Foundation of China Grants No.~12075101, No.~12235016.

\appendix
\section{Green's function on torus}\label{A}
In this appendix, we give an elementary derivation of Green's functions on a torus with modular parameter $\tau$. The Green's functions $G(z,w)$ and $\tilde{G}(z,w)$ satisfy the defining equation
\begin{align}
    \frac{1}{\pi}\partial_{\bar{z}} G(z,w) = \delta(z,w) - \frac{1}{\text{Im}\tau}
\end{align}
and
\begin{align}
    \frac{1}{\pi}\partial_z\partial_{\bar{z}} \tilde{G}(z,w) = \delta(z,w) - \frac{1}{\text{Im}\tau}
\end{align}
where $\delta(z,w)$ is the delta function with respect to the measure $d^2 z = \frac{i}{2}dz\wedge d\bar{z}$. The Green's functions can be rewritten as $G(z-w)$ and $\tilde{G}(z-w)$ by translational invariance.

For the Green's function $G(z)$, since it takes the form $\frac{1}{z}$ in the complex plane, it is tempting to represent the Green's function on the torus as a formal series
\begin{align}
 \sum_{(m,n)\in \mathbb{Z}^2} \frac{1}{z-(m+n\tau)}   
\end{align}
with manifest double periodicity. To make the formal series convergent, it's natural to add the holomorphic terms $\sum\limits_{(m,n)\in \mathbb{Z}^2 \backslash (0,0)} \frac{1}{(m+n\tau)} - \frac{z}{(m+n\tau)^2}$ and we obtain
\begin{align}
    \zeta_\tau(z) = \frac{1}{z} + \sum_{(m,n)\in \mathbb{Z}^2 \backslash (0,0)} \Big(\frac{1}{z-(m+n\tau)} + \frac{1}{(m+n\tau)} + \frac{z}{(m+n\tau)^2}\Big)
\end{align}
which is known as the Weierstrass Zeta function. However, the double periodicity is broken by the added terms. We have
\begin{align}
    &\zeta_\tau(z+1) - \zeta_\tau(z) \nonumber\\
    = &\lim_{M\to \infty} \Big\lbrace \frac{1}{z+1} + \sum_{(m,n)\in \mathbb{Z}^2 \backslash (0,0),|m|\leq M} \Big(\frac{1}{z+1-(m+n\tau)} + \frac{1}{(m+n\tau)} + \frac{z+1}{(m+n\tau)^2}\Big) \nonumber\\
    &- \Big[ \frac{1}{z} + \sum_{(m,n)\in \mathbb{Z}^2 \backslash (0,0),|m|\leq M} \Big(\frac{1}{z-(m+n\tau)} + \frac{1}{(m+n\tau)} + \frac{z}{(m+n\tau)^2}\Big) \Big] \Big\rbrace \nonumber\\
    = &\lim_{M \to \infty} \Big\lbrace \sum_{n\in \mathbb{Z}} \Big(\frac{1}{z+M+1-n\tau} - \frac{1}{z-M-n\tau}\Big) +  \sum_{(m,n)\in \mathbb{Z}^2 \backslash (0,0),|m|\leq M} \frac{1}{(m+n\tau)^2} \Big\rbrace \nonumber\\
    = &\lim_{M \to \infty} \Big\lbrace \sum_{n\in \mathbb{Z}} \Big(\frac{1}{M+1-n\tau} - \frac{1}{-M-n\tau}\Big) +  \sum_{(m,n)\in \mathbb{Z}^2 \backslash (0,0),|m|\leq M} \frac{1}{(m+n\tau)^2} \Big\rbrace
\end{align}
so $\zeta_\tau(z+1) - \zeta_\tau(z)$ is independent on $z$. Since $\zeta_\tau(z)$ is an odd function, we have
\begin{align}
    \zeta_\tau(z+1) - \zeta_\tau(z) = \zeta_\tau(\frac{1}{2}) - \zeta_\tau(-\frac{1}{2}) = 2 \zeta_\tau(\frac{1}{2})
\end{align}
and similarly
\begin{align}
    \zeta_\tau(z+\tau) - \zeta_\tau(z) = \zeta_\tau(\frac{\tau}{2}) - \zeta_\tau(-\frac{\tau}{2}) = 2 \zeta_\tau(\frac{\tau}{2})
\end{align}
Therefore we can restore the double periodicity by adding a linear function, obtaining Green's function on the torus
\begin{align}
    G(z) = \zeta_\tau (z) + \frac{2}{\tau-\bar{\tau}} [(\bar{\tau}\zeta_\tau(\frac{1}{2}) - \zeta_\tau(\frac{\tau}{2})) z + (-\tau\zeta_\tau(\frac{1}{2}) + \zeta_\tau(\frac{\tau}{2})) \bar{z}]
\end{align}
Acting by $\partial_{\bar{z}}$ we get the defining equation
\begin{align}
    \frac{1}{\pi}\partial_{\bar{z}} G(z) = \delta(z) + \frac{2}{\pi(\tau-\bar{\tau})}(-\tau\zeta_\tau(\frac{1}{2}) + \zeta_\tau(\frac{\tau}{2}))
\end{align}
We must have $\tau\zeta_\tau(\frac{1}{2}) - \zeta_\tau(\frac{\tau}{2}) = i\pi$, then the Green's function can be further simplified to
\begin{align}
    G(z) = \zeta_\tau (z) - 2\zeta_\tau(\frac{1}{2}) z + \frac{2\pi i}{\text{Im}\tau}\text{Im}z
\end{align}

Similarly, the Green's function $\tilde{G}(z)$ takes the form $\log(z\bar{z})$ in the complex plane, it's tempting to represent the Green's function on the torus as a formal series with explicit double periodicity
\begin{align}
    \sum_{(m,n)\in \mathbb{Z}^2} \log[(z-(m+n\tau))(\bar{z}-(m+n\bar{\tau}))]
\end{align}
We can add harmonic terms to obtain a convergent series
\begin{align}
    \log(|\sigma_\tau(z)|^2) = &\log(z\bar{z}) + \sum_{(m,n)\in \mathbb{Z}^2\backslash (0,0)} [\log(\frac{z-(m+n\tau)}{m+n\tau}\frac{\bar{z}-(m+n\bar{\tau})}{m+n\bar{\tau}}) \notag\\
    &+ \frac{z}{m+n\tau} + \frac{z^2}{2(m+n\tau)^2} + \frac{\bar{z}}{m+n\bar{\tau}} + \frac{\bar{z}^2}{2(m+n\bar{\tau})^2}]
\end{align}
where
\begin{align}
    \sigma_\tau(z) = z \prod_{(m,n)\in \mathbb{Z}^2\backslash (0,0)} (1-\frac{z}{m+n\tau}) e^{\frac{z}{m+n\tau} + \frac{z^2}{2(m+n\tau)^2}}
\end{align}
is the Weierstrass sigma function, with its log derivative being the Weierstrass zeta function. We have
\begin{align}
    &\partial_z (\log(|\sigma_\tau(z+1)|^2) - \log(|\sigma_\tau(z)|^2)) = \zeta_\tau(z+1) - \zeta_\tau(z) = 2\zeta_\tau(\frac{1}{2}) \notag\\
    &\partial_{\bar{z}} (\log(|\sigma_\tau(z+1)|^2) - \log(|\sigma_\tau(z)|^2)) = \zeta_{\bar{\tau}}(\bar{z}+1) - \zeta_{\bar{\tau}}(\bar{z}) = 2\overline{\zeta_\tau(\frac{1}{2})}
\end{align}
so
\begin{align}
    \log(|\sigma_\tau(z+1)|^2) - \log(|\sigma_\tau(z)|^2) = 2\zeta_\tau(\frac{1}{2})z + 2\overline{\zeta_\tau(\frac{1}{2})}\bar{z} + \zeta_\tau(\frac{1}{2}) + \overline{\zeta_\tau(\frac{1}{2})}
\end{align}
knowing that $\log(|\sigma_\tau(z)|^2)$ is an even function. Similarly, we have
\begin{align}
    \log(|\sigma_\tau(z+\tau)|^2) - \log(|\sigma_\tau(z)|^2) = 2\zeta_\tau(\frac{\tau}{2})z + 2\overline{\zeta_\tau(\frac{\tau}{2})}\bar{z} + \tau \zeta_\tau(\frac{\tau}{2}) + \bar{\tau} \overline{\zeta_\tau(\frac{\tau}{2})}
\end{align}
Therefore, we can restore the double periodicity by adding a quadratic function
\begin{align}
    \tilde{G}(z) &= \log(|\sigma_\tau(z)|^2) - (\zeta_\tau(\frac{1}{2}) - \frac{\pi}{2\text{Im}\tau}) z^2 - (\overline{\zeta_\tau(\frac{1}{2})} - \frac{\pi}{2\text{Im}\tau}) \bar{z}^2 - \frac{\pi}{\text{Im}\tau} z\bar{z} \notag\\
    &= \log(|\sigma_\tau(z)|^2) - \zeta_\tau(\frac{1}{2}) z^2 - \overline{\zeta_\tau(\frac{1}{2})} \bar{z}^2 - \frac{2 \pi}{\text{Im}\tau} (\text{Im}z)^2
\end{align}
and it's easy to verify $\tilde{G}(z)$ satisfies the defining equation.

\section{Global regularity condition}\label{B}
In section \ref{2.1} we determine the perturbed one-point correlator up to a constant by the Weyl anomaly and conservation equation. The constant should be determined by the condition that the metric in the Fefferman-Graham gauge in $\rho\in(0,\infty)$ can be extended to a global solution including $\rho=0$ after a diffeomorphism. As a mnemonic, we call the condition the global regularity condition. To the first order the change by diffeomorphism is given by the Lie derivative of the unperturbed metric in the direction of the vector $V^{[1]}$, which vanishes on the conformal boundary and is assumed to be sufficiently differentiable in $\rho \in (0,+\infty)$. Now to the first order, the perturbed bulk metric is given by
\begin{align} \label{Perturbed metric first order in epsilon}
    ds'^2 =& ds^2 + \epsilon g^{FG[1]}_{ij} dx^i dx^j + \epsilon {\cal L}_{V^{[1]}} ds^2 +O(\epsilon^2) 
\end{align}
where $g^{FG[1]}$ is given by the first order variation of $g^{(0)}$ and $g^{(2)}$
\begin{align} \label{delta gFG1}
    g^{FG[1]}_{zz} &= g^{(2)[1]}_{zz} - \frac{e^{-2\rho}}{\tau^2} g^{(2)[1]}_{z\bar{z}} + e^{2\rho} 
    g^{(0)[1]}_{zz} - \frac{e^{-2\rho}}{\tau^4} g^{(0)[1]}_{\bar{z}\bar{z}} \nonumber\\
    g^{FG[1]}_{z\bar{z}} &= -\frac{e^{-2\rho}}{2 \bar{\tau}^2} g^{(2)[1]}_{zz} -\frac{e^{-2\rho}}{2 \tau^2}  g^{(2)[1]}_{\bar{z}\bar{z}} + g^{(2)[1]}_{z\bar{z}} + (e^{2\rho} - \frac{e^{-2\rho}}{\tau^2 \bar{\tau}^2}) g^{(0)[1]}_{z\bar{z}} \nonumber\\
   g^{FG[1]}_{\bar{z}\bar{z}} &= g^{(2)[1]}_{\bar{z}\bar{z}} - \frac{e^{-2\rho}}{\bar{\tau}^2} g^{(2)[1]}_{z\bar{z}} + e^{2\rho} g^{(0)[1]}_{\bar{z}\bar{z}} - \frac{e^{-2\rho}}{\bar{\tau}^4} g^{(0)[1]}_{zz}
\end{align}
which in turn are determined from (\ref{holographic reconstruction g0 first order}) and (\ref{holographic reconstruction g2 first order}). The components of the Lie derivative of the unperturbed metric are 
\begin{align} \label{Lie derivative of metric}
    & {\cal L}_{V^{[1]}} g_{\mu\nu} \nonumber\\
    =& l^2
    \begin{pmatrix}
        2 \partial_\rho {V^{[1]}}^\rho & -\frac{\pi^2}{\tau^2} \partial_\rho {V^{[1]}}^z + \frac{\pi^2}{2} (e^{2\rho} + \frac{e^{-2\rho}}{\tau^2\bar{\tau}^2}) \partial_\rho {V^{[1]}}^{\bar{z}} & \frac{\pi^2}{2} (e^{2\rho} + \frac{e^{-2\rho}}{\tau^2\bar{\tau}^2}) \partial_\rho {V^{[1]}}^z - \frac{\pi^2}{\bar{\tau}^2} \partial_\rho {V^{[1]}}^{\bar{z}} \\
        .. & 0 & \pi^2 (e^{2\rho} -\frac{e^{-2\rho}}{\tau^2\bar{\tau}^2}) {V^{[1]}}^\rho \\
        .. & .. & 0
    \end{pmatrix}\nonumber \\
    & +\partial_z 
    \begin{pmatrix}
        ...
    \end{pmatrix}_{3\times3} 
    +\partial_{\bar{z}} 
    \begin{pmatrix}
        ...
    \end{pmatrix}_{3\times3},  
\end{align}
where we omit the symmetric parts and the total derivatives with respect to $\partial_z$ and $\partial_{\bar{z}}$.
\footnote{We will see later that the final form of regularity is the integration (\ref{regularity 1''''}) and (\ref{regularity 2''''}) over the torus, and the total derivative terms can be omitted because their contributions vanish when integrated.}

The metric is regular at the center circle $\rho = \rho_0$, if its components in $(\phi,x,y)$ coordinates, which properly covers $\rho = \rho_0$, are regular. Let $g_{\mu\nu}$ and $g_{\tilde{\mu}\tilde{\nu}}$ be the components of the metric in $(\rho,z,\bar{z})$ and $(\phi,x,y)$ respectively. We have the transformation of the components
\begin{align}
    g_{\mu\nu} &= g_{\tilde{\alpha}\tilde{\beta}} \frac{\partial x^{\tilde{\alpha}} }{\partial x^\mu} \frac{\partial x^{\tilde{\beta}} }{\partial x^\nu}, 
\end{align}
where the transformation matrix is 
\begin{align}
    (\frac{\partial x^{\tilde{\alpha}} }{\partial x^\mu})=
    \begin{pmatrix}
        0 & \frac{2\pi \bar{\tau}}{\bar{\tau}-\tau} & \frac{2\pi \tau}{\tau- \bar{\tau}} \\
        \cos[\pi(\frac{z}{\tau} + \frac{\bar{z}}{\bar{\tau}})] & -\frac{\pi (\rho-\rho_0)}{\tau} \sin[\pi(\frac{z}{\tau} + \frac{\bar{z}}{\bar{\tau}})] & -\frac{\pi (\rho-\rho_0)}{\bar{\tau}} \sin[\pi(\frac{z}{\tau} + \frac{\bar{z}}{\bar{\tau}})]\\
        \sin[\pi(\frac{z}{\tau} + \frac{\bar{z}}{\bar{\tau}})] & \frac{\pi (\rho-\rho_0)}{\tau} \cos[\pi(\frac{z}{\tau} + \frac{\bar{z}}{\bar{\tau}})] & \frac{\pi (\rho-\rho_0)}{\tau} \sin[\pi(\frac{z}{\tau} + \frac{\bar{z}}{\bar{\tau}})]
    \end{pmatrix}.
\end{align}
Taking the $\rho\to \rho_0$ limit, for a regular metric $g$, the $(\phi,x,y)$-components $g_{\tilde{\mu}\tilde{\nu}}$ should go to a finite limit. This can thus give us the behavior of the $(\rho,z,\bar{z})$-components $g_{\mu\nu}$ near $\rho \sim \rho_0$. Consider the components $g_{\rho\rho}$, $\tau^2 g_{zz} + \bar{\tau}^2 g_{\bar{z}\bar{z}} + 2\tau \bar{\tau} g_{z\bar{z}}$ and $\tau^2 g_{zz} - \bar{\tau}^2 g_{\bar{z}\bar{z}}$, expand them with respect to $\epsilon$, and then we take the first order terms, 
\begin{gather}
    g_{\rho\rho}^{[1]}=\frac{1}{2}\big(g_{xx}^{[1]} +g_{yy}^{[1]} +(g_{xx}^{[1]}-g_{yy}^{[1]})\cos[2\pi (\frac{z}{\tau} +\frac{\bar{z}}{\bar{\tau}})] +2g_{xy}^{[1]} \sin[2\pi (\frac{z}{\tau} +\frac{\bar{z}}{\bar{\tau}})] \big), \\
    \tau^2 g_{zz}^{[1]} - \bar{\tau}^2 g_{\bar{z}\bar{z}}^{[1]} = \frac{8\pi^2 \tau \bar{\tau} (\rho-\rho_0)}{\tau-\bar{\tau}} (g_{\phi x}^{[1]} \sin[\pi (\frac{z}{\tau} +\frac{\bar{z}}{\bar{\tau}})] -g_{\phi y}^{[1]} \cos[\pi (\frac{z}{\tau} +\frac{\bar{z}}{\bar{\tau}})]), \\
    \tau^2 g_{zz}^{[1]} + \bar{\tau}^2 g_{\bar{z}\bar{z}}^{[1]} + 2\tau \bar{\tau} g_{z\bar{z}}^{[1]}=2\pi^2(\rho-\rho_0)^2 \big(g_{xx}^{[1]} +g_{yy}^{[1]} +(g_{yy}^{[1]}-g_{xx}^{[1]}) \cos[2\pi (\frac{z}{\tau} +\frac{\bar{z}}{\bar{\tau}})]  \notag \\
    -2g_{xy}^{[1]} \sin[2\pi (\frac{z}{\tau} +\frac{\bar{z}}{\bar{\tau}})] \big).
\end{gather}
On the other hand, from (\ref{Perturbed metric first order in epsilon}) we know that 
\begin{align}
    &g_{\rho\rho}^{[1]}= {\cal L}_{V^{[1]}} g_{\rho\rho}, \\
    &\tau^2 g_{zz}^{[1]} - \bar{\tau}^2 g_{\bar{z}\bar{z}}^{[1]} = \tau^2 g_{zz}^{FG[1]} - \bar{\tau}^2 g_{\bar{z}\bar{z}}^{FG[1]}, \\
    &\tau^2 g_{zz}^{[1]} + \bar{\tau}^2 g_{\bar{z}\bar{z}}^{[1]} + 2\tau \bar{\tau} g_{z\bar{z}}^{[1]} = \tau^2 g_{zz}^{FG[1]} + \bar{\tau}^2 g_{\bar{z}\bar{z}}^{FG[1]} + 2\tau \bar{\tau} g_{z\bar{z}}^{FG[1]} + 2\tau \bar{\tau} {\cal L}_{V^{[1]}} g_{z\bar{z}} .
\end{align}
where we omit the total derivative with respect to $\partial_z$ and $\partial_{\bar{z}}$. We equate the right-hand sides of these two sets of equations and then get three equations.
Note that the term $(g_{xx} +g_{yy})$ in the third equation can be eliminated by the first equation,
\begin{equation}
    g_{xx} +g_{yy}= 2 {\cal L}_{V^{[1]}} g_{\rho\rho} - (g_{xx}-g_{yy})\cos[2\pi (\frac{z}{\tau} +\frac{\bar{z}}{\bar{\tau}})] -2g_{xy} \sin[2\pi (\frac{z}{\tau} +\frac{\bar{z}}{\bar{\tau}})].
\end{equation}
Then from the second and the third equation, we have
\begin{align}
    & g_{\phi x}^{[1]} \sin[\pi (\frac{z}{\tau} +\frac{\bar{z}}{\bar{\tau}})] -g_{\phi y}^{[1]} \cos[\pi (\frac{z}{\tau} +\frac{\bar{z}}{\bar{\tau}})]= \frac{\tau-\bar{\tau}}{8\pi^2 \tau \bar{\tau} (\rho-\rho_0)} ( \tau^2 g_{zz}^{FG[1]} - \bar{\tau}^2 g_{\bar{z}\bar{z}}^{FG[1]} ), \label{regularity 1}\\
    & (g_{yy}^{[1]} -g_{xx}^{[1]}) \cos[2\pi (\frac{z}{\tau} +\frac{\bar{z}}{\bar{\tau}})] -2g_{xy}^{[1]} \sin[2\pi (\frac{z}{\tau} +\frac{\bar{z}}{\bar{\tau}})]= \nonumber\\
    &\qquad \frac{1}{4\pi^2(\rho-\rho_0)^2} ( \tau^2 g_{zz}^{FG[1]} + \bar{\tau}^2 g_{\bar{z}\bar{z}}^{FG[1]} + 2\tau \bar{\tau} g_{z\bar{z}}^{FG[1]} +2\tau \bar{\tau} {\cal L}_{V^{[1]}} g_{z\bar{z}}) - {\cal L}_{V^{[1]}} g_{\rho\rho}. \label{regularity 2}
\end{align}
The regularity of the metric lead to the fact that the left-hand sides of (\ref{regularity 1}) and (\ref{regularity 2}) have finite value at $\rho=\rho_0$. Expanding (\ref{regularity 1}) and (\ref{regularity 2}) with respect to $(\rho-\rho_0)$, the coefficient of (\ref{regularity 1}) in order of $(\rho-\rho_0)^{-1}$ is
\begin{equation}
    \frac{\tau-\bar{\tau}}{8\pi^2 \tau \bar{\tau}} ( \tau^2 g_{zz}^{FG[1]} - \bar{\tau}^2 g_{\bar{z}\bar{z}}^{FG[1]} )\big|_{\rho=\rho_0} =0 ,\label{regularity 1'}
\end{equation}
and the coefficient of (\ref{regularity 2}) in order of $(\rho-\rho_0)^{0}$ is
\begin{align}
    & (g_{yy}^{[1]} -g_{xx}^{[1]})\big|_{\rho=\rho_0} \cos[2\pi (\frac{z}{\tau} +\frac{\bar{z}}{\bar{\tau}})] -2g_{xy}^{[1]}\big|_{\rho=\rho_0} \sin[2\pi (\frac{z}{\tau} +\frac{\bar{z}}{\bar{\tau}})]= \nonumber\\
    &\quad \frac{1}{8\pi^2} \partial_\rho^2 ( \tau^2 g_{zz}^{FG[1]} + \bar{\tau}^2 g_{\bar{z}\bar{z}}^{FG[1]} + 2\tau \bar{\tau} g_{z\bar{z}}^{FG[1]} +2\tau \bar{\tau} {\cal L}_{V^{[1]}} g_{z\bar{z}})\big|_{\rho=\rho_0} - {\cal L}_{V^{[1]}} g_{\rho\rho}\big|_{\rho=\rho_0}. \label{regularity 2'}
\end{align}
Note that from (\ref{Lie derivative of metric}) we find 
\begin{equation}
    \frac{\tau \bar{\tau}}{4\pi^2} \partial_\rho^2 {\cal L}_{V^{[1]}} g_{z\bar{z}} \big|_{\rho=\rho_0} = {\cal L}_{V^{[1]}} g_{\rho\rho}\big|_{\rho=\rho_0},
\end{equation}
so (\ref{regularity 2'}) can be simplified to
\begin{multline}
    (g_{yy}^{[1]} -g_{xx}^{[1]})\big|_{\rho=\rho_0} \cos[2\pi (\frac{z}{\tau} +\frac{\bar{z}}{\bar{\tau}})] -2g_{xy}^{[1]}\big|_{\rho=\rho_0} \sin[2\pi (\frac{z}{\tau} +\frac{\bar{z}}{\bar{\tau}})]= \\
    \quad \frac{1}{8\pi^2} \partial_\rho^2 ( \tau^2 g_{zz}^{FG[1]} + \bar{\tau}^2 g_{\bar{z}\bar{z}}^{FG[1]} + 2\tau \bar{\tau} g_{z\bar{z}}^{FG[1]} )\big|_{\rho=\rho_0} .\label{regularity 2''}
\end{multline}
Integrate (\ref{regularity 2''}) over a period of $(z,\bar{z})$ and we get 
\begin{align}
    &\int_{\text{T}^2} d^2 z \partial_\rho^2 ( \tau^2 g_{zz}^{FG[1]} + \bar{\tau}^2 g_{\bar{z}\bar{z}}^{FG[1]} + 2\tau \bar{\tau} g_{z\bar{z}}^{FG[1]} )\big|_{\rho=\rho_0} =0, \label{regularity 2'''}
\end{align}
since the integration of the trigonometric functions on the left-hand side of (\ref{regularity 2''}) vanishes.

Plug (\ref{delta gFG1}) into (\ref{regularity 1'}) and (\ref{regularity 2'''}), and we obtain two equations on the integration of ${g^{(2)[1]}}_{zz}$ and ${g^{(2)[1]}}_{\bar{z}\bar{z}}$ over $\text{T}^2$. The solution is
\begin{align}
    \int_{\text{T}^2} d^2 z g^{(2)[1]}_{zz} =& \frac{2\pi^2 \bar{\tau}}{\tau^3} \int_{\text{T}^2} d^2 z f_{\bar{z}\bar{z}},\label{regularity 1''''}\\
    \int_{\text{T}^2} d^2 z g^{(2)[1]}_{\bar{z}\bar{z}} =& \frac{2\pi^2 \tau}{\bar{\tau}^3} \int_{\text{T}^2} d^2 z f_{zz},\label{regularity 2''''}
\end{align}
where (\ref{T1 trace eqn}) and (\ref{holographic reconstruction g2 first order}) are used to simplify the results.



\bibliographystyle{JHEP}
\bibliography{reference.bib}

\end{document}